\documentclass[preprint]{aastex}

\slugcomment{Accepted by ApJ: Jan 28, 2000}
\shortauthors{van der Tak et~al. }
\shorttitle{Structure and evolution of massive YSO envelopes}

\usepackage{psfig}

\def\hii{\ion{H}{2}}
\def\htwo{H$_2$}
\def\co{$^{13}$CO}
\def\cs{C$^{34}$S}
\def\water{H$_2$O}

\def\hhco{H$_2$CO}

\def\gtsim{{_>\atop{^\sim}}}
\def\ltsim{{_<\atop{^\sim}}}
\def\vlsr{V$_{\rm LSR}$}

\def\tbol{T$_{\rm bol}$}
\def\kms{km~s$^{-1}$}

\def\scm{cm$^{-2}$}
\def\ccm{cm$^{-3}$}
\def\mic{$\mu$m}
\def\klm{k$\lambda$}
\def\mass{$M(<r_0)$}

\def\msol{M$_{\odot}$}
\def\lsol{L$_{\odot}$}

\begin{document}

\title{ Structure and Evolution of the
  Envelopes of Deeply Embedded Massive Young Stars}

\author{Floris F. S. van der Tak \& Ewine F. van Dishoeck}
\affil{Sterrewacht, Postbus 9513, 2300 RA Leiden, The Netherlands}
\authoraddr{Sterrewacht, Postbus 9513, 2300 RA Leiden, The Netherlands}
\author{Neal J. Evans II} 
\affil{Department of Astronomy, University of Texas, Austin, TX 78712}
\and
\author{Geoffrey A. Blake}
\affil{Division of Geological and Planetary Sciences, California
Institute of Technology, MS 150--21, Pasadena, CA 91125} 

\begin{abstract}
  
  The physical structure of the envelopes around a sample of fourteen
  massive young stars is investigated using maps and spectra in
  submillimeter continuum and lines of C$^{17}$O, CS, \cs\ and \hhco.
  Nine of the sources are highly embedded luminous ($10^3-10^5$~\lsol)
  young stellar objects which are bright near-infrared sources, but
  weak in radio continuum; the other objects are similar but not
  bright in the near-infrared, and contain ``hot core''-type objects
  and/or ultracompact \ion{H}{2} regions.  The data are used to
  constrain the temperature and density structure of the circumstellar
  envelopes on $10^2-10^5$~AU scales, to investigate the relation
  between the different objects and to search for evolutionary
  effects.
 
  The total column densities and the temperature profiles are obtained
  by fitting self-consistent dust models to submillimeter photometry.
  The calculated temperatures range from 300~to 1000~K at $\sim
  10^2$~AU and from 10~to 30~K at $\sim 10^5$~AU from the star. Visual
  extinctions are a few hundred to a few thousand magnitudes, assuming
  a grain opacity at $\lambda 1300$\mic, of $\approx 1$~\scm~g$^{-1}$
  of dust, as derived earlier for one of our sources. The mid-infrared
  data are consistent with a $30$\% decrease of the opacity at higher
  temperatures, caused by the evaporation of the ice mantles.
  
  The CS, \cs\ and \hhco\ data as well as the submillimeter dust
  emission maps indicate density gradients $n\propto r^{-\alpha}$.
  Assuming a constant CS abundance throughout the envelope, values of
  $\alpha=1.0-1.5$ are found, significantly flatter than the $\alpha =
  2.0 \pm 0.3$ generally found for low-mass objects. This flattening
  may indicate that in massive young stellar objects, nonthermal
  pressure is more important for the support against gravitational
  collapse, while thermal pressure dominates for low-mass sources. We
  find $\alpha=2$ for two hot core-type sources, but regard this as an
  upper limit since in these objects, the CS abundance may be enhanced
  in the warm gas close to the star.  
  
  The assumption of spherical symmetry is tested by modeling infrared
  absorption line data of \co, CS emission line profiles and
  near-infrared continuum. There is a distinct, but small deviation
  from spherical symmetry: the data are consistent with a decrease of
  the optical depth by a factor of $\approx 3$ in the central $\ltsim
  10''$. The homogeneity of the envelopes is verified by the good
  agreement of the total masses in the power law models with the
  virial masses.  
  
  Modeling of C$^{17}$O emission shows that $\approx 40-90$\% of the
  CO is frozen out onto the dust. The CO abundances show a clear
  correlation with temperature, as expected if the abundance is
  controlled by freeze-out and thermal desorption. The CS abundance is
  $3\times 10^{-9}$ on average, ranging from $(4-8)\times 10^{-10}$ in
  the cold source GL~7009S to $(1-2)\times 10^{-8}$ in the two ``hot
  core''-type sources.
  
  Dense outflowing gas is seen in the CS and H$_2$CO line wings; the
  predominance of blueshifted emission suggests the presence of dense,
  optically thick material within 10$''$ of the center.
  Interferometric continuum observations at $\lambda 1300-3500$\mic\ 
  show compact emission, probably from an $0\farcs3$ diameter,
  optically thick dust component, such as a dense shell or a disk. The
  emission is a factor of $10-100$ stronger than expected for the
  envelopes seen in the single-dish data, so that this component may
  be opaque enough to explain the asymmetric CS and \hhco\ line
  profiles.
  
  The evolution of the sources is traced by the overall temperature
  (measured by the far-infrared color), which increases
  systematically with decreasing ratio of envelope mass to stellar
  mass. The observed anticorrelation of near-infrared and radio
  continuum emission suggests that the erosion of the envelope
  proceeds from the inside out. Conventional tracers of the evolution
  of low-mass objects do not change much over this narrow age range. 

\end{abstract}


\section{Introduction}
\label{s:intr}

The dynamical processes governing massive star formation are much less
understood than is the case for low-mass stars \citep{chur99,garay99}.
The various observational ``appearances'' of low-mass star formation
(T~Tauri stars, FU~Orionis stars, infrared Class I-III sources,
molecular outflows, ...)  have been linked into a single, if rough,
general evolutionary scenario \citep{shu93,andr93,andr99,evans99}.  In
contrast, no clear evolutionary sequence has been established for
high-mass stars. Objects such as the BN~object in Orion are highly
embedded and emit the bulk of their luminosity in the mid- and
far-infrared.  Ultracompact (UC) \ion{H}{2} regions are small ($\sim
0.1$~pc) sources of free-free emission at radio wavelengths
\citep{wood89a,kurt94}.  ``Hot core''-type objects have bright
molecular line emission at submillimeter wavelengths, which indicate
temperatures of several $100$~K and high abundances of saturated
carbon-bearing molecules (e.g., Blake et al.\ 1987).  They usually
lack detectable radio emission, which might be due to ``quenching'' by
material accreting at rates $\gtsim 10^{-7}$~\msol~yr$^{-1}$
\citep{walm95b}.  The distinction between UC~\ion{H}{2} regions and
hot cores is not always clear, and the two are often found located
very close to each other \citep{cesar99,kurt99}.  As a step towards
understanding the evolution of massive YSOs, this paper presents
models for a sample of fourteen such objects.

The formation of high-mass stars is invisible at optical wavelengths
because of the high opacity of the surrounding material, so that
reliable age estimates for massive protostellar objects are difficult
to obtain. Such estimates for massive protostars have traditionally
come from the morphology of the radio continuum emission
\citep{coll80}, but this method applies only to relatively evolved
objects. More recently, the size to velocity ratio or dynamical time
scale of molecular outflows has been used as a measure of age, but the
derived age depends strongly on the adopted dynamical model
\citep{cabr97}. The chemical composition of the material is a
potentially powerful clock, but difficult to calibrate
\citep{char92,evdgab98}.

The density and temperature structure of the circumstellar material
are also expected to change as the central star develops. For the
density, power laws $n(r) \propto r^{-\alpha}$ are predicted by many
theoretical models.  Before collapse begins, low-mass cores may relax
to a power law, with $\alpha=2.0$ for thermal support \citep{shu77},
and $\alpha=1.0$ for turbulent support
\citep{lizan89,myer92,mclaug96}.  Once collapse begins, the density
distribution tends toward $\alpha=1.5$ \citep{lars69, shu77}.  The
temperature structure $T(r)$ is determined by the luminosity, the dust
opacities, and $n(r)$.  Since the response of $T(r)$ to changes in
luminosity is rapid, we use the observed luminosity to determine
$T(r)$.  Besides giving information on the dynamics of star formation,
a good model of the physical structure of objects is a prerequisite
for determining their chemical composition, which then by itself can
be used as a powerful additional evolutionary indicator.  The physical
conditions around newly-born massive stars also reveal some of the
influence that such stars have on their surroundings, which is of
interest for the effects of star formation on large ($\gtsim$~1~pc)
scales.

The physical structure of massive YSO envelopes has been studied based
on dust continuum observations by, e.g., \citet{chin86},
\citet{cww90}, \citet{fais98} and \citet{cesar98}.  
Such studies are sensitive to the column density and the temperature
of the envelope, but not to density itself. A major source of
uncertainty for all dust models is the choice of optical properties of
the grains, which limits the accuracy of derived envelope masses to a
factor of 2 \citep{henn97a}.

Molecular rotational lines are direct probes of the \htwo\ density and
are also sensitive to temperature.  \citet{zhou91,zhou94} and
\citet{carr95} investigated the structure of massive star-forming
regions using lines of CS, \hhco\ and other molecules with large
dipole moments. However, in these studies, the column density (or the
cloud mass) was not independently constrained. Doing this requires
knowledge of the molecular abundance, information which is generally
unavailable except in the case of CO, which is relatively inert and
much more abundant than most molecules. Still, part of the CO may be
frozen out onto grain surfaces in the cold regions far from
embedded stars.

This paper uses observations of both the dust and the molecular gas to
constrain the physical structure of the envelopes of a sample of
massive young stars.  The study builds on the method of analysis
developed in a previous paper for one source, GL 2591, by
\citet{fvdt99}.  The column density is inferred from continuum
observations, and the density from molecular lines, while the
temperature is calculated self-consistently from the luminosity, dust
properties, and $n(r)$. We use the derived density and temperature
profiles to characterize the early evolution of massive stars and
their surroundings.

\section{Choice of Targets}
\label{s:samp}

The fourteen sources studied in this paper are introduced in Table~1,
which lists the source names, positions, distances, luminosities,
radio continuum flux densities and associated IRAS Point Source
Catalog (PSC) entries. All sources have been well studied before at
radio and infrared wavelengths.  They were selected to be luminous
($>10^3$~\lsol) and visible from the Northern hemisphere.
The distances to some of the sources are quite uncertain; in
particular, for the sources in the Cygnus~X region, we use a fiducial
distance of 1~kpc \citep{fvdt99}, for ease of scaling once better
distances are available.

The main sample consists of nine deeply embedded massive young stars,
which were additionally selected for mid-infrared brightness
($>100$~Jy at $\lambda 12$~\mic), allowing comparison to existing
absorption line studies.  A combined analysis of emission and
absorption lines gives powerful constraints on the physical structure
and geometry of the system, and allows a nearly complete inventory of
the chemical composition of both the solid and the gas phases. These
nine sources have been extensively studied in the infrared, both from
the ground \citep{will82,mitc90} and with the Infrared Space
Observatory \citep{evd97}. As a comparison sample, five luminous
embedded young stars which are weak in the mid-infrared are studied,
whose surroundings contain a ``hot core'' and/or an ultracompact
\ion{H}{2} region.

The full sample contains sources with luminosities between $1\times
10^3$~and $2\times 10^5$~\lsol\ and distances between $0.9$~and
$4$~kpc. The radio flux densities are at the $\sim$mJy level for most
sources, which is orders of magnitude lower than expected from an
\ion{H}{2} region photoionized by a star with the same luminosity.
The radio emission from the bright mid-infrared sources has a spectral
index $\gamma$ in the range $0.5-1.0$ and may arise in a spherical
wind for which $\gamma =0.7$.  The principal difference between our
main sample and that of other studies of massive young stars (e.g.,
\citealt{tofan95,hunt98,srid99}) is the selection on mid-infrared
brightness, which could introduce an orientation bias, in the sense
that low-density cavities such as produced by molecular outflows may
be preferentially directed towards us. Alternatively, the high
mid-infrared brightness may be an evolutionary effect. Depending on
the relative importance of different effects, the mid-infrared bright
sources in the main sample could be younger or older than the sources
in the comparison sample. We will discuss this point in more detail in
Section~\ref{sec:evol}.

The sources were selected on isolated location within $\sim 30''$ at
infrared wavelengths, so that the heating is dominated by the central
star.  However, while GL~2591 and GL~490 are examples of objects
forming in isolation, NGC~6334, S~140, NGC~7538 and W~3 are regions
where low-mass stars are also born, as revealed by near-infrared
images \citep{straw89,evan89,hodap94,megeat96,bloom98}.  In addition,
S~140 contains three massive objects separated by $10-15''$, which
contribute about equally to the far-infrared luminosity
\citep{evan89}. For the other sources, this information is not
available.  In the case of W~3~IRS5, radio continuum observations
\citep{claus94} suggest the presence of seven early B-type stars
within $4''$, a very large number in view of the average mass function
and stellar density in young clusters \citep{carp93,luhm98,alves98}.
As long as the power source is much smaller than our beam, its
multiplicity does not matter for our analysis.

In several cases, high-resolution continuum observations reveal the
presence of an (ultra-)compact \hii\ region next to the source of
interest. Since luminosity and/or mass are dominated by the infrared
source, our models of centrally heated sources are still reasonable
approximations. 

\section{Observations}
\label{s:obs}

\subsection{Single Dish Submillimeter Spectroscopy}

Spectroscopy of selected molecular lines in the $230$, $345$ and
$490$~GHz bands was performed with the 15-m James Clerk Maxwell
Telescope (JCMT)\footnote{The James Clerk Maxwell Telescope is
  operated by the Joint Astronomy Centre, on behalf of the Particle
  Physics and Astronomy Research Council of the United Kingdom, the
  Netherlands Organization for Scientific Research and the National
  Research Council of Canada.} on Mauna Kea, Hawaii during various
runs in 1995--1997. In total, 15-20 lines were observed per source.
The antenna has an approximately Gaussian main beam of FWHM $18''$ at
230 GHz, $14''$ at 345~GHz, and $11''$ at 490 GHz. Detailed
technical information about the JCMT and its receivers and
spectrometer can be found on-line at {\tt
  $<$http://www.jach.hawaii.edu/JCMT/home.html$>$}. Receivers A2, B3i
and C2 were used as front ends at 230, 345 and 490 GHz, respectively.
The Digital Autocorrelation Spectrometer served as backend, with
continuous calibration and natural weighting employed. Values
for the main beam efficiency, $\eta_{mb}$, determined by the JCMT staff
from observations of Mars and Jupiter, were $0.69$, $0.58$ and $0.53$
at 230, 345 and 490~GHz for the 1995 data, and $0.64$, $0.60$ and
$0.53$ for 1996 and 1997. Absolute calibration should be correct to
$30\%$, except for data in the 230~GHz band from May 1996, which have
an uncertainty of $\approx 50\%$ due to technical problems with
receiver A2. Pointing was checked every 2 hours during the observing
and was always found to be within $5''$. Integration times are 30--40
minutes per frequency setting, resulting in rms noise levels in
$T_{\rm mb}$ per $625$~kHz channel ranging from $\approx 30$~mK at
230~GHz to $\approx 100$~mK at 490~GHz.

To subtract the atmospheric and instrumental background, a reference
position $180''$~East was observed using the chopping secondary
mirror. For the C$^{17}$O $J=2\to 1$ line, we also position-switched
using an $1800''$ offset, which increased the line fluxes typically by
15\%.  The contribution by extended emission in other lines should be
less since they all need higher densities for the excitation.  Where
two measurements of C$^{17}$O\ $J=2\to1$ are available, the
position-switched data will be used.

Observations of the \cs\ $J=2\to 1$ and $J=3\to 2$ and CS $J=5\to 4$
lines were carried out with the 30-m telescope of the Institut de
Radio Astronomie Millim\'etrique (IRAM) on Pico Veleta, Spain, on
January 28-30, 1999. Receivers B100, C150 and B230 were used at 96,
145 and 245~GHz, respectively, tuned single sideband. The
autocorrelator was used as the backend, which covers a bandwidth of
$110-170$~\kms\ at a resolution of $0.2$~\kms.  The FWHM beam sizes
are $24''$, $16''$ and $10.4''$; the forward efficiencies were $0.90$,
$0.82$ and $0.84$ and the beam efficiencies $0.75$, $0.55$ and $0.48$.
The data were calibrated onto the $T_{\rm mb}$ scale by multiplying
$T^*_A$ by the ratio of the forward and beam efficiencies. 
As a calibration check, observations of GL~2591 and
NGC~7538~IRS1 were compared to results from \citet{plum97} and found
to agree to $10$\%. 

The data on GL~2591 have been presented in \citet{fvdt99}. The
observations on GL~490 are given in Schreyer et al. (2000, in prep.).
Most of the data on W~3~(\water) and W~3~IRS5 are taken from
\citet{helm97}. In addition, we used data from the surveys by
\citet{angl96}, \citet{bron96}, \citet{plum92,plum97}, and from the
observations of \citet{kast94} for GL~2136, \citet{zhou94} for
S140~IRS1, \citet{cesar97} for IRAS~20126, \citet{haus93} and
\citet{mang93} for DR~21~(OH), Dartois (1998, Ph.D.\ thesis) for
GL~7009S and Olmi \& Cesaroni (1999, A\&A, in press) for W~28~A2. Care
was taken not to include data at positions $>5''$ away from those in
Table~\ref{sam.tab}. For the sources GL~7009S, IRAS 20126+4104 and
W~28~A2, no \hhco\ data are available.

 Line parameters were measured by fitting a
Gaussian profile, where the free parameters were the total line flux
$\int T_{\rm MB}dV$, the FWHM line width and the center velocity. The
measured line fluxes are collected in Table~\ref{t:lineflux};
Table~\ref{vlsr.tab} gives for each source the central LSR velocity
and the FWHM line width. The values are the averages of the C$^{17}$O
and \cs\ lines (4~lines in most cases), which are assumed to be
optically thin.  The error bars reflect the spread among the lines,
which for all sources exceeds the error in the individual measurements
for both quantities.  None of the CS and H$_2$CO line profiles show
obvious evidence of self-absorption.

\subsection{Single-dish Mapping}
\label{s:obs_maps}

Maps of the CS $J=5\to4$ and/or $7\to6$ and/or C$^{18}$O $J=2\to1$
lines were obtained in 1998 July and December and 1999 July with the
$10.4$-m antenna of the Caltech Submillimeter Observatory
(CSO)\footnote{The Caltech Submillimeter Observatory is operated by
the California Institute of Technology under funding from the U.S.
National Science Foundation (AST96-15025)}.  The backends were the
facility 500 and 50 MHz bandwidth Acousto-Optical Spectrometers (AOS).
The pointing is accurate to within $4''$. Beam sizes and main beam
efficiencies were $32''$ and 0.66 at 230~GHz and $21''$ and 0.61 at
345~GHz.

In July 1998, submillimeter continuum maps at $\lambda350$~\mic\ of
W~33A, GL~2136, S~140, GL~490, GL~2591 and W~3~IRS5 were made using
the Submillimeter High Angular Resolution Camera (SHARC)
\citep{hunt96} on the CSO.  The CSO has a beam of size $10''$ at this
wavelength.  The weather was good, with zenith optical depths at
$\lambda350$~\mic\ of $0.92-1.66$. From observations of Uranus, the
gain was measured to be $(140 \pm 30)$ Jy/V.

\subsection{Interferometer Observations}
\label{sec:obs_ovro}

Maps at $86-230$~GHz of GL~2136, NGC~7538~IRS1, NGC~7538~IRS9 and
W~33A were obtained with the six-element interferometer of the Owens
Valley Radio Observatory (OVRO)\footnote{The Owens Valley Millimeter
  Array is operated by the California Institute of Technology under
  funding from the U.S. National Science Foundation (AST96-13717).}.
The OVRO interferometer consists of six 10.4~m antennas on North-South
and East-West baselines. A detailed technical description of the
instrument is given in \citet{padin91}. Data were collected in
1997-1999 in several array configurations at frequencies of $86$,
$106$ and $115$~GHz. The two sources in NGC~7538 were observed in the
compact and extended array configurations, while for the Southern sources
GL~2136 and W~33A, a hybrid configuration with long North-South and
short East-West baselines was also used to improve the beam
shape. Baseline lengths range from the shadowing limit out to $\sim
80$~\klm\ at $86$~GHz and out to $150$~\klm\ at $230$~GHz,
corresponding to spatial frequencies of $\sim 2500$ to $\sim
10^5$~rad$^{-1}$. 

The antenna gains and phases were monitored with short integrations of
nearby quasars: BL~Lac for NGC~7538~IRS1 and IRS9, and NRAO~530 for
W~33A and GL~2136. Passband corrections were derived using a local
signal generator and by fitting first order polynomials to data on
3C273. Flux calibration is based on snapshots of Uranus and Neptune,
and is estimated to be accurate to $\approx 30$~\% at $86-115$~GHz and
to $\approx 50$~\% at $230$~GHz.

Simultaneous continuum observations in the $230$~GHz window were also
performed, but only in the cases of W~33A and NGC~7538~IRS1 did the
weather allow useable data to be taken. In addition, observations of
molecular lines at $86-230$~GHz were carried out with the OVRO
digital correlator, the results of which will be presented elsewhere.

\section{Results}
\label{s:res}

\subsection{Molecular Emission Line Profiles}
\label{s:res:spectra}

\begin{figure}[hp]
  
  \psfig{figure=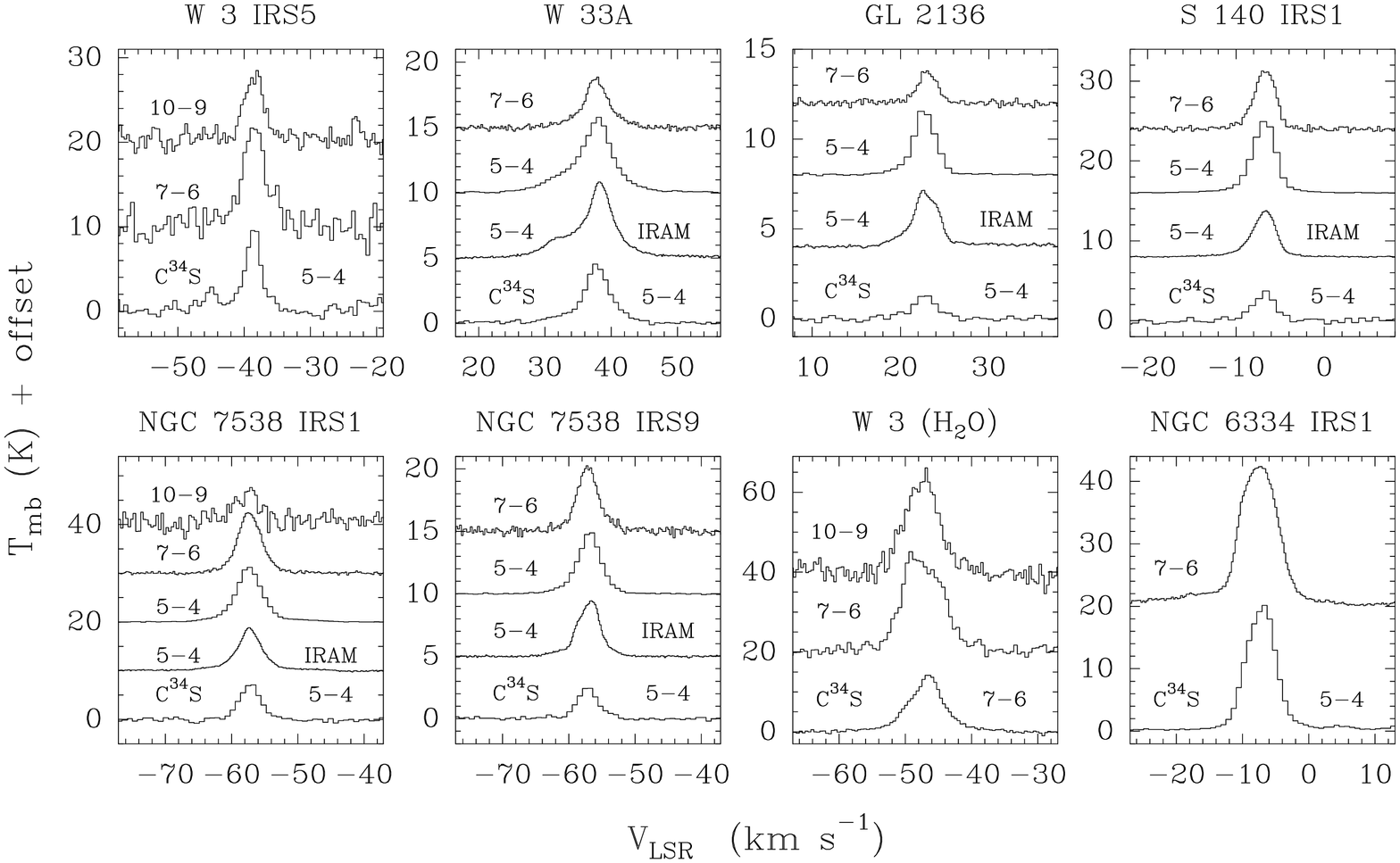,angle=-90,width=17cm}

  \caption{Line profiles of CS and \cs, observed 
    with the JCMT and IRAM~30m telescopes. For clarity, the IRAM data
    have been divided by~2 and the \cs\ data have been multiplied by 2
    for the hot cores and by 3 for the other sources. At blueshifted
    velocities, wings are visible on the CS lines, most prominently in
    the IRAM data.}
  \label{f:jcmt_res}
\end{figure}

The CS line profiles are presented in Figure~\ref{f:jcmt_res}. In
addition to a strong, single-peaked line core, which is also detected
in isotopic lines (\cs, \co) and which has an approximately Gaussian
shape, high-velocity wings are clearly detected. These wings are more
prominent in $J=5\to 4$ than in higher-$J$ lines, are also detected in
the strongest \hhco\ lines, and must arise on small scales, since they
are stronger in the IRAM~30m beam than in the JCMT beam.  The wings
are not detected in the W~3 sources, for which we do not have $J=5\to
4$ data.  Since the sources studied in this paper are known to drive
CO outflows, it is natural to associate the CS and \hhco\ wings with
dense gas in these outflows.

In all bright mid-infrared sources studied in this paper except
W~33A, blueshifted infrared absorption of CO is detected at similar
velocities as in CO rotational line emission \citep{mitc91a}, and in
the cases of W~3~IRS5 and GL~2591 up to much higher velocities,
implying that the outflow is directed along the line of sight.  In
realistic outflow models with a finite opening angle, a mixture of
blue- and redshifted emission is expected on both sides of the driving
source, which should be visible in submillimeter emission.  However,
the wings seen on the CS and \hhco\ lines are much stronger at
blueshifted than at redshifted velocities, and for some sources no
redshifted wing emission is detected at all. Since the line asymmetry
is stronger in the IRAM~30m spectrum of CS $J=5\to 4$ than in the JCMT
profile of the same line, the asymmetry must arise on scales of
$\ltsim 10''$. Since the redshifted outflow lobe lies in the
background, we suggest that the lack of redshifted CO and CS wing
emission is due to obscuration. Since the outflow lies at a velocity
offset, the absorption must be continuous absorption by dust. The
\htwo\ column density required to do this is $\sim 10^{25}$~\scm\ on a
$\ltsim 10''$ scale, corresponding to a visual extinction of $\sim
10^4$ magnitudes.

\subsection{Average Physical Conditions}
\label{s:est_phys}

Estimates of the mean temperature and density of the gas can be
obtained by comparing observed line ratios of specific molecules with
non-LTE models which include radiative trapping.  Examples of this
approach can be found in \citet{jans94} and \citet{helm94}; it is
similar to the ``Large Velocity Gradient'' method used by
\citet{zhou94} and \citet{plum97}. The observed line ratios of \cs\
are well-suited to constrain the \htwo\ density.  We have calculated
synthetic line ratios for a range of temperatures and densities, using
the rate coefficients by S.~Green (priv.\ comm., cf.\
\citealt{turn92}). Comparison with the data indicates densities of
$10^5$ to $>10^7$~\ccm, but different line ratios generally give
inconsistent answers for the same source, indicating the presence of 
density inhomogeneities, probably in the form of a gradient since
lines observed with several telescopes are systematically brighter
when observed with a smaller beam.

The measured ratios of the \hhco\ $J=5_{24}\to 4_{23}$/$J=5_{05}\to
4_{04}$ and the $J=5_{42/41}\to 4_{41/40}$/$J=5_{24}\to 4_{23}$ lines
are particularly sensitive to temperature. Unlike \cs, these \hhco\
line ratios were measured in the same beam, or even in the same
spectrum, improving the calibration between pairs of lines, although
their filling factors might still be different. Model calculations
using collisional rate coefficients by \citet{green91} give
temperatures of $\approx 60$~K for S~140~IRS1 to $>200$~K for
W~3~(\water) and NGC~6334~IRS1. Again, the two ratios usually do not
yield the same temperature for the same source, indicating the need
for models with a varying temperature.  In general, the modeling
indicates somewhat higher temperatures and densities for the weak
mid-infrared sources. In Section~\ref{s:models}, we will see that this
effect is due to a steeper density gradient in these sources.

\begin{figure}[hp]
  \begin{center}
    
  \psfig{figure=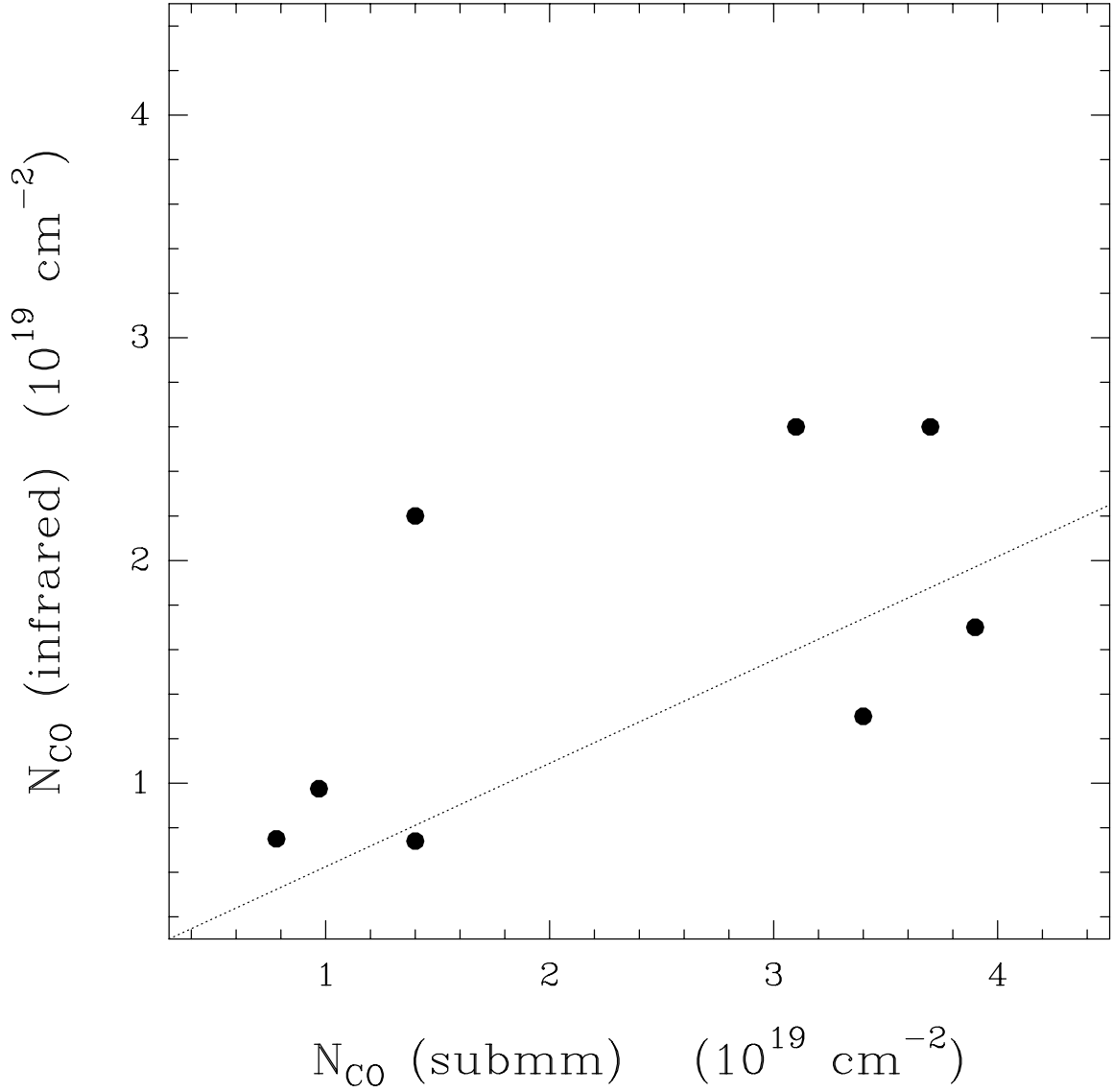,width=17cm}

    \caption{Column densities of CO derived from C$^{17}$O emission
    compared to values derived from infrared 
      absorption of \co. The dotted line indicates the relation expected 
      for a constant-density model.}

    \label{f:coldens}
  \end{center}
\end{figure}

The CO column densities derived from C$^{17}$O, using the temperatures
and densities found above, are listed in column~3 of
Table~\ref{vlsr.tab}.  A plot of these values against the column
densities derived by \citet{mitc90} from infrared absorption
observations is shown in Figure~\ref{f:coldens}. Abundance ratios of
$^{12}$C/$^{13}$C = 60 and $^{16}$O/$^{17}$O = 2500 \citep{wils94} are
assumed. The two measurements agree to a factor of~2 for all sources,
and often much better. This result implies that the circumstellar
envelopes of our objects have to first order a spherical shape, with
the infrared source in the center (see also Section~\ref{sec:struc}).
The average absorption column density is $66$\% of the emission value,
which is higher than half as expected in a uniform model. Beam
dilution in the emission data could explain this difference, in which
case the sources are centrally condensed.

\subsection{Submillimeter Continuum and Line Maps}
\label{s:res_maps}

\begin{figure}[hp]

 \vskip-5cm  
 \psfig{figure=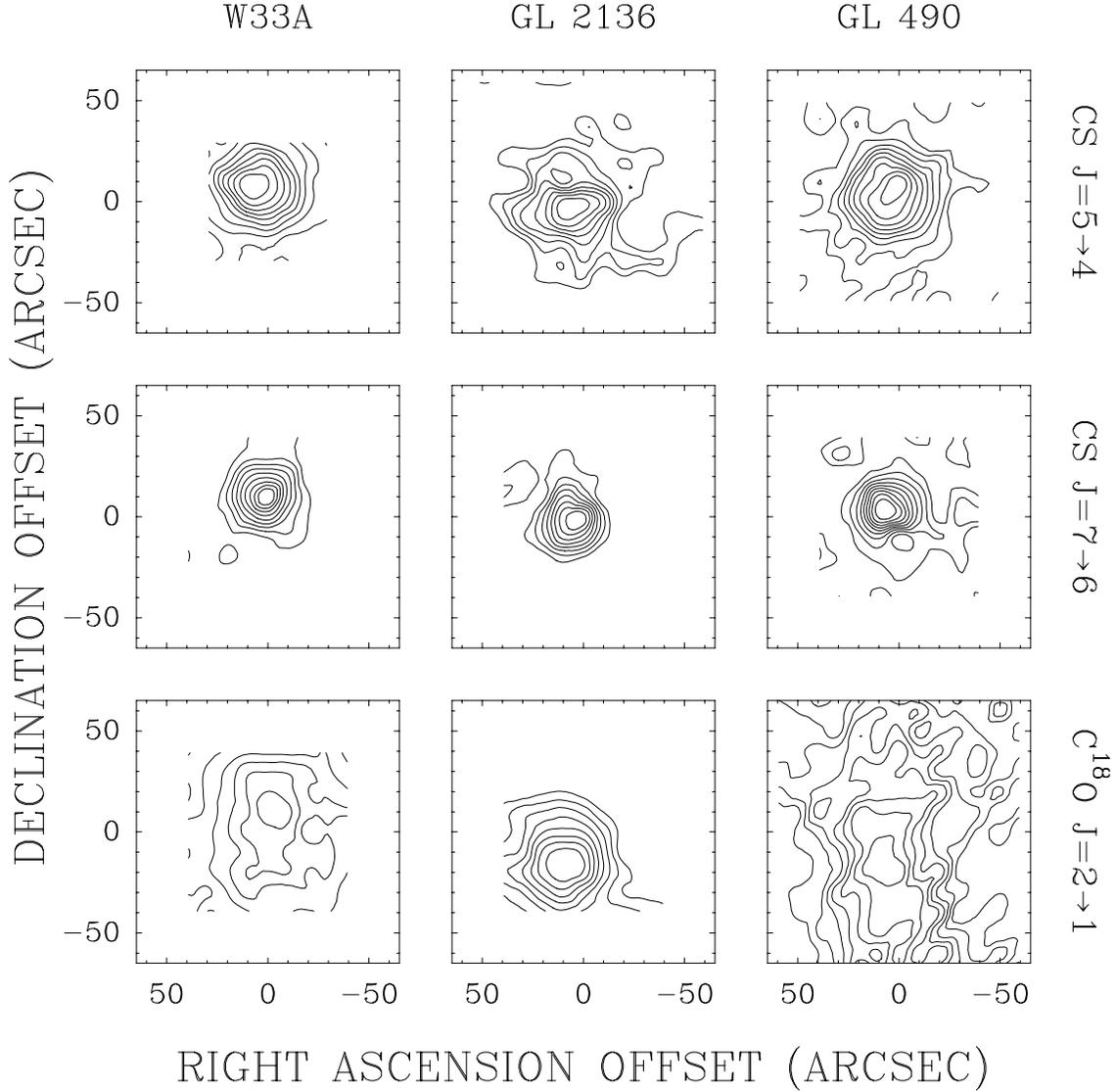,width=15cm}

  \caption{CSO maps of the C$^{18}$O $J=2\to 1$ and CS $J=5\to 4$ and
    $7\to 6$ lines. Contours are 10\% of the peak flux, starting at
    10\% for CS and 30\% for C$^{18}$O. Integration intervals and peak
    C$^{18}$O fluxes are: 19 to 26~\kms\ and 3.5~K~\kms\ for GL~2136,
    -15 to -10 \kms\ and 4.0~K~\kms\ for GL~490, -18 to 0 \kms\ and
    7.7~K~\kms\ for NGC~6334, -65 to -50 \kms\ and 5.0~K~\kms\ for
    NGC~7538, -12 to -2 \kms\ and 3.5~K~\kms\ for S~140, 30 to 44
    \kms\ and 3.8~K~\kms\ for W~33A, -58 to -38 \kms\ and 5.1~K~\kms\
    for W~3 (\water) and -50 to -32 \kms\ and 5.2~K~\kms\ for W~3
    (Main). See Table~\ref{t:lineflux} for the peak fluxes of the CS
    lines.}  \label{f:lmaps}

\end{figure}

The CS and C$^{18}$O maps are presented in Figure~\ref{f:lmaps};
Figure~\ref{f:cmaps} shows the SHARC $\lambda 350$~\mic\ maps and maps
of NGC~7538 at $\lambda\lambda 450,850$~\mic, obtained with the
Submillimeter Common-User Bolometer Array (SCUBA) on the JCMT,
provided by G.~Sandell (1998, private communication).

The maps appear compact but slightly extended. In a few cases, the map
peak is offset from the center position, but this offset falls within
the $5''$ pointing uncertainty. In the case of CS, this suggests that
the infrared sources coincide with local density maxima in the
surrounding molecular cloud. The C$^{18}$O maps are sometimes peaked
at the same position as CS, which implies a maximum of the column
density, but more often, these maps are not strongly peaked and have a
much lower dynamic range than the CS maps. The map of NGC~7538 shows a 
chaotic structure, rather than clear peaks.

Column~4 of Table~\ref{vlsr.tab} lists the diameter of the CS $J=5\to
4$ emission, measured from the maps as the point where the brightness
has dropped to 50\% of the maximum. These numbers will be used in the
next section to constrain the radii of the models. The values span a
fairly narrow range: $36-64''$ and do not show an obvious correlation
with distance, suggesting that the cores do not have a single
intrinsic linear size.  Instead, it appears that the molecular gas has
a scale-free density structure, such as can be described by a power
law.  This conclusion is supported by the fact that the CS diameters
are somewhat ($10-100$\%) larger than the beam size, which is also
what power law models predict.

The maps do not show a cutoff in the emission, such as would be
produced if the star-forming cores had a distinct edge. Rather, the
brightness keeps going down until the detection limit. Maps of a wider
field and with a higher dynamic range than presented here may reveal
if there is such an edge or if at large radii, the density gradient
flattens out and the core merges into a surrounding molecular cloud.
Some objects, such as W~3~IRS5, DR~21 (OH), NGC~6334 IRS1 and the
NGC~7538 sources, are clearly part of a more extended giant molecular
cloud. However, the focus of this paper is on the $\sim 30''$ region
around the massive young stars.

\subsection{Interferometric Continuum Maps}
\label{sec:res_ovro}

\begin{figure}[hp]
  \begin{center}
    
  \psfig{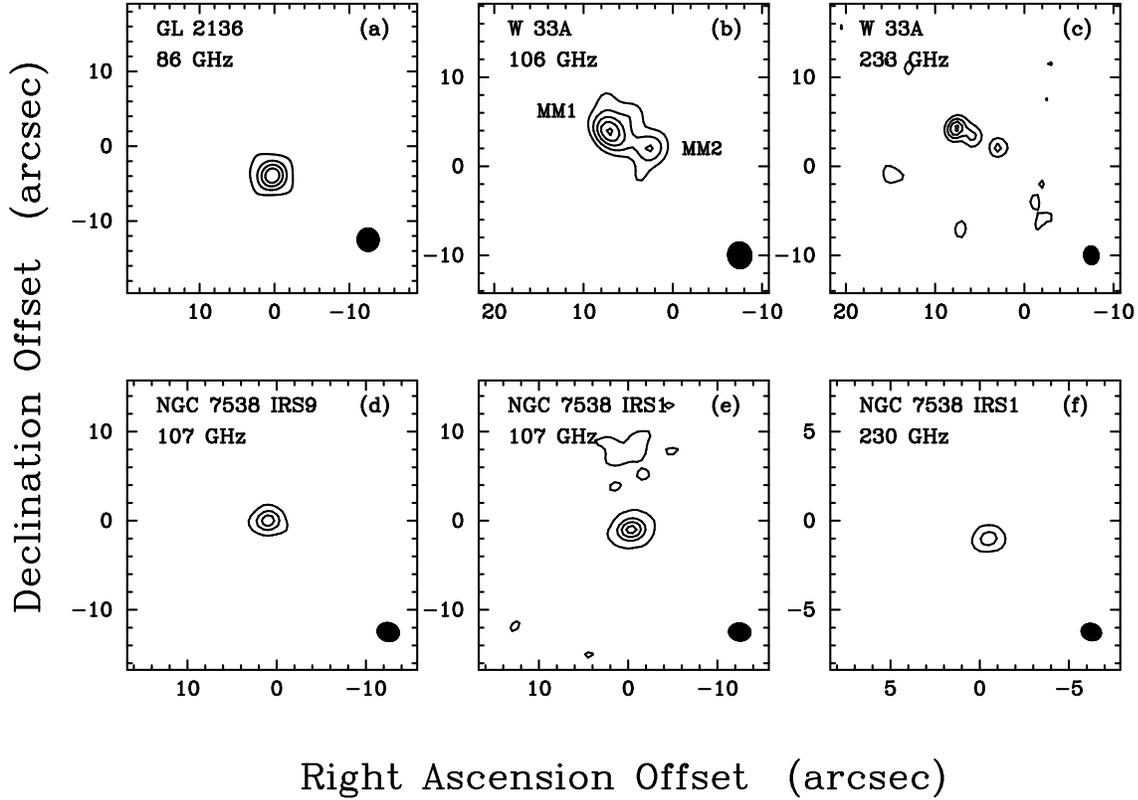}

    \caption{OVRO continuum maps of GL 2136, W 33A, NGC 7538 IRS1 and
      NGC 7538 IRS9. Contour levels (in mJy/beam) are: 6, 18, 30, 42
      (a); 8, 16, 24, 32, 40, 48, 56 (b), 40, 80, 120, 160, 200, 240,
      280 (c); 6, 18, 30, 42 (d), 50, 550, 1050, 1550 (e), 500, 2000,
      3500, 5000, 6500, 8000 (f). The beam sizes and shapes are drawn
      in the bottom right corner of each panel.}
    \label{fig:ovromaps}
  \end{center}
\end{figure}

Figure~\ref{fig:ovromaps} presents maps created from the OVRO data by
gridding and Fourier transforming the visibility data with uniform
weighting.  Deconvolution with the CLEAN algorithm and
self-calibration of the $uv$ phases based on the brightest CLEAN
components helped to improve the image quality.
Table~\ref{posflux.tab} lists the positions and flux densities of the
detected sources.

The maps typically show a single compact source at or near phase
center, within the $\sim 5''$ positional error of the infrared
positions from Table~\ref{sam.tab}. The extended $107$~GHz emission
North of NGC~7538~IRS1 is the \hii\ region surrounding NGC~7538~IRS2,
which has also been detected at lower frequencies with the VLA (e.g.,
\citealt{henk84}).  In the case of W~33A, two sources are detected,
neither of which coincides with the IR position of \citet{dyck77}.
The brightest source, MM1, coincides however with a VLA detection
\citep{reng96} and with the infrared position by \citet{capp78}, who
also report a second $2.2$~\mic\ source $3''$~South of W~33A, which
may be related to MM2. It is unknown if these two sources are
physically associated; if they are, they may represent a young massive
binary star at a separation of $4\farcs7$ or $19,000$~AU. This
separation is larger than the $\sim10^3$~AU found by \citet{wyrow99b}
for W~3~(\water) and by \citet{mund99} for a sample of low-mass young
stellar objects, but comparable to that found by \citet{wood89} and
\citet{padin89} for DR~21~(OH).

The flux densities found here for NGC~7538~IRS1 are in agreement with
those by \citet{wood89} and \citet{akab92}. The spectral index,
measured from 107 to 230 GHz, is $(0.9\pm 0.5)$.  The spectral indices
of the sources in W~33A are $(1.8\pm 0.6)$ (MM1) and $(1.6\pm 0.6)$
(MM2).  The values of the spectral indices for W~33A rule out
optically thin emission from either ionized gas or from dust. They
are, however, consistent with black body emission, probably from a
compact dust source. In Paper~1, OVRO observations of GL~2591 were
presented which gave similar results; we refer the reader to that
paper for a detailed discussion.

\section{Models}
\label{s:models}

Motivated by the results of the excitation analysis
(\S~\ref{s:est_phys}) and of the CS maps (\S~\ref{s:res_maps}), we
will proceed by developing spherically symmetric power law models.
The modeling procedure follows \citet{fvdt99} to which we refer for
details.

\subsection{Dust Continuum Models: Mass and Temperature Structure}
\label{s:mod_dust}

The dust emission from the sources was modeled using the
one-dimensional diffusion code by \citet{egan88}. A density structure
of the form $n=n_0 (r/r_0)^{-\alpha}$ was used, with $\alpha$ in the
range $0.5-2.0$.
The fiducial radius $r_0$ was set to be half the diameter of the CS
emission ($D(CS)$ in Table~\ref{vlsr.tab}), converted to AU in
Table~\ref{t:m_gas}.
The density $n_0$ at this radius was derived for each $\alpha$ by
matching the submillimeter dust emission; in the next section, we will
constrain the value of $\alpha$ by modeling molecular lines.  To avoid
edge effects in modeling the full range of emission, we used outer
radii in our models that are twice $r_0$.  The inner radius was
arbitrarily set to 1/300 of the outer radius, small enough not to
influence the calculated brightness or temperature profile, as
verified by test calculations for W~33A and GL~2591. Dust opacities
appropriate for star-forming regions were taken from \citet{ossen94}
(their Model~5) and converted to absorption cross-sections using a
grain mass density of $2.5$~g~\ccm\ and a radius of $0.1$~\mic. This
radius is close to the median of more realistic size distributions, so
that the calculated dust temperatures represent the bulk of the dust
(cf.\ \citealt{cww90}). Dust properties are known to vary from one
region to the next \citep{lis98,viss98,mrh00} and are likely to change
inside our sources as well, due to the changing temperature
(\citealt{menn98}; see \S~\ref{s:ice}). Maps at several far-infrared
wavelengths at $\ltsim 10''$ resolution would allow us to disentangle
temperature and dust opacity variations, but awaiting such data, we
assume a constant grain opacity.  With the luminosity, the distance
and the size of the source fixed, \mass\ was used as the only free
parameter to match the observed submillimeter continuum emission.

\begin{figure}[hp]
  
  \psfig{figure=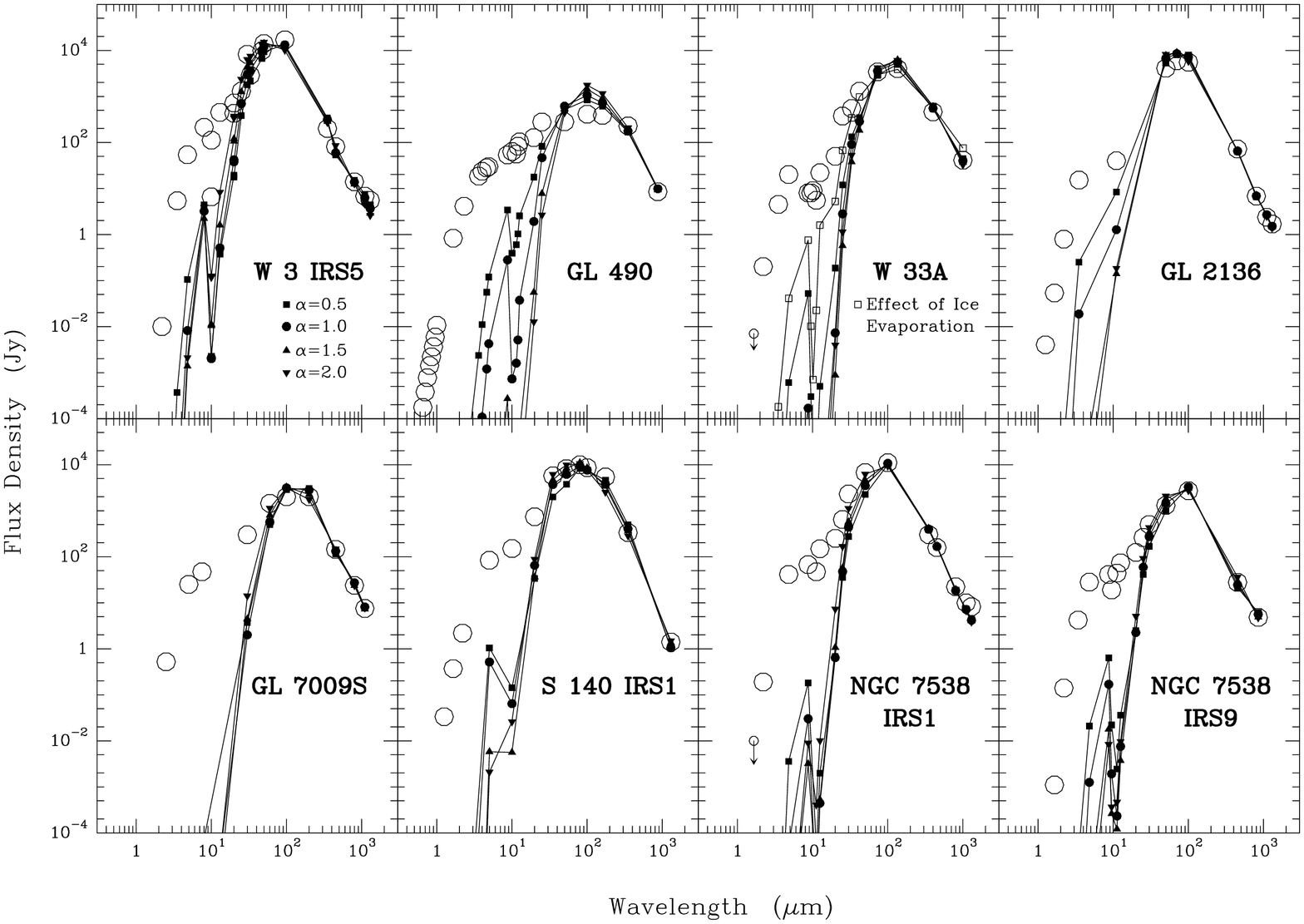,angle=-90,width=17cm}

  \caption{}
    Observed continuum spectra ({\it open circles}) and models 
    for several values of $\alpha$. The symbol legend is in the top
    left panel. For W~33A, a model using grains without ice mantles is
    also shown. Data are taken from \citet{camp95} for W~3~IRS5;
     \citet{chin91} for GL~490; \citet{moor81}, \citet{evan79},
     \citet{dyck77}, \citet{stier84}, \citet{jaff84} and
     \citet{cheun80} for W~33A; \citet{allen77}, \citet{lebof76}, van
     Dishoeck (priv.~comm.) and \citet{kast94} for GL~2136;
     \citet{dart98b} and \citet{mccut95} for GL~7009S; \citet{evan89},
     \citet{will82} and \citet{zhou94} for S~140~IRS1, \citet{wern79}
     for NGC~7538 IRS1; \citet{wern79} and G.~Sandell (priv.~comm.) for
     NGC~7538 IRS9; \citet{wilk90} and \citet{cesar97,cesar99} for IRAS
     20126+4104; \citet{thum75}, \citet{harv86,harv77}, \citet{rich89a}
     and \citet{chan93b} for DR~21~(OH); \citet{wynn72}, \citet{keto92}
     and \citet{thron79} for W~3~(\water), \citet{straw89} and
     \citet{harv83} for NGC~6334~IRS1; and \citet{moor81},
     \citet{ligh84} and \citet{harv94} for W~28~A2.

  \label{f:dusto}
\end{figure}

Figure~\ref{f:dusto} shows the synthetic continuum spectra compared
with observations. Most submillimeter data are from this work and from
\citet{sand94}; the sources of the additional data are listed in the
caption. The models have been constructed to fit the submillimeter
($\lambda 300-1300$\mic) data, and are seen to reproduce the observed
emission at $\lambda \gtsim 50$~\mic\ for every source, i.e.\ up to
optical depths of a few. The shorter-wavelength emission is in general
not matched, although care was taken to include only photometry in
small ($\ltsim 10''$) beams to avoid a contribution from reflection
nebulosities. The high brightness of GL~2591 at $\lambda 20$\mic\ was
attributed by \citet{fvdt99} to the evaporation of ice mantles close
to the star, which decreases the $\lambda 20$\mic\ optical depth by
30\% \citep{ossen94}.  Similar effects play a role for the sources
presented here, as illustrated for W~33A in the figure. The emission
at $\lambda \ltsim 10$\mic\ is not expected to be well reproduced in
these models, since the high optical depth makes it very sensitive to
deviations from the assumed spherical shape. The fact that all sources
require less extinction at short wavelengths is discussed
quantitatively in \S~\ref{sec:struc}. This point is illustrated by the
fact that the column densities derived in this paper are much higher
than those found by \citet{fais98} by fitting the near- to
far-infrared spectra of a sample of ultracompact \ion{H}{2} regions,
which have similar submillimeter flux densities and lie at similar
distances.

The calculated temperatures follow an $r^{-0.4}$ profile at distances
greater than $\sim 2-3 \times 10^3$~AU from the star, with the
absolute temperature scale set by the luminosity. Inside this radius,
the envelope is optically thick to the photons carrying most of the
energy, and the temperature gradient is steeper than $r^{-0.4}$, with
the absolute scale set by the extinction, which acts as a blanket.
Hence, for a given luminosity, a lower extinction leads to a higher
temperature at a certain radius. This is the case for our models with
shallow density gradients.  For a given column density, a shallower
density gradient implies a lower extinction (in a pencil beam),
because a larger fraction of the beam is filled with warm dust. We
therefore expect, like \citet{chin86} and \citet{cww90}, that bright
mid-infrared sources have shallow density gradients.  However, it is
seen from Figure~\ref{f:dusto} that even the models with $\alpha=0.5$
fail to fit the near-infrared part of the spectrum.  This is not due
to our selection on brightness in the mid-infrared since the data of
the comparison samples are not matched either. The reason for the
discrepancy must be sought in (small) geometrical effects.

\subsection{Models of the Line Emission: Density Structure}
\label{s:mod_gas}

To determine the slope of the density gradient $\alpha$, the
C$^{17}$O, CS, \cs\ and \hhco\ line radiation from the~14 sources was
modeled with the Monte Carlo code written by \citet{hst00}.  The
models consist of~30 spherical shells spaced logarithmically within
the same radii as the dust models. The same density structures were
used as in the dust models, and the temperature structure was taken
from the dust continuum models for the same value of $\alpha$. The
intrinsic line profile was assumed to be a Gaussian with the measured
FWHM (Table~\ref{vlsr.tab}). Initially, molecular abundances were
assumed to be constant throughout the model.

The gas column density (or equivalently \mass\ or the value of $n_0$)
for each $\alpha$ was also taken from the dust models, assuming a gas
to dust mass ratio of 100. This assumption was tested for GL~2591
against C$^{17}$O observations by \citet{fvdt99} because this source
has negligible depletion of CO (see also \S~\ref{s:ice}). With the
column density fixed, the density profile can be obtained by modeling
the emission lines of the high-critical density molecules CS and
\hhco. After solving for the molecular excitation, velocity-integrated
intensities have been calculated in the appropriate beam for each
observation. In many cases, a line was observed in several beams,
which in our models acts as a simple substitute for simulating map
data.  Comparison to the observations proceeds by minimizing the
quantity
$$ \chi^2 \equiv \frac{1}{N} \Sigma \left( \frac{F^{\rm (obs)} - F^{\rm
(model)})}{\sigma}\right)^2, $$
where $F^{\rm (obs)}$ and $F^{\rm (model)}$ are the observed and
    synthetic line fluxes and the sum is taken over all $N$ observed
    lines.  A grid of models was run for each molecule by varying the
    density law exponent $\alpha$ and the molecular abundance.

The success of any fit procedure depends critically on a good
understanding of the error budget ($\sigma$). Most line fluxes were
assigned a 30\% error, but the following lines have a larger
uncertainty of 50\%: {(i)} lines in the 460-490~GHz band, because of
the more uncertain calibration; { (ii)} lines measured in beams of
$\gtsim 30''$ diameter, which may suffer from confusion; and {(iii)}
the $J=1\to 0$ lines of CS and \cs, which may contain a contribution
from the surrounding cloud. The following observations
were found to be more than a factor of 2 off any of our models, and
therefore discarded in the $\chi^2$ calculation: CS $J=5\to4$ and
$7\to6$ in GL~490 and CS $7\to6$ in W~33A from \citet{plum97}, CS
$5\to4$ and \cs\ $3\to2$ in S140~IRS1 from \citet{zhou94}, \cs\ 
$2\to1$ in IRAS 20126+4104 from \citet{cesar97} and \cs\ $J=7\to 6$ in
W~28~A2 from \citet{plum97}.

\begin{figure}[hp]
    
   \psfig{figure=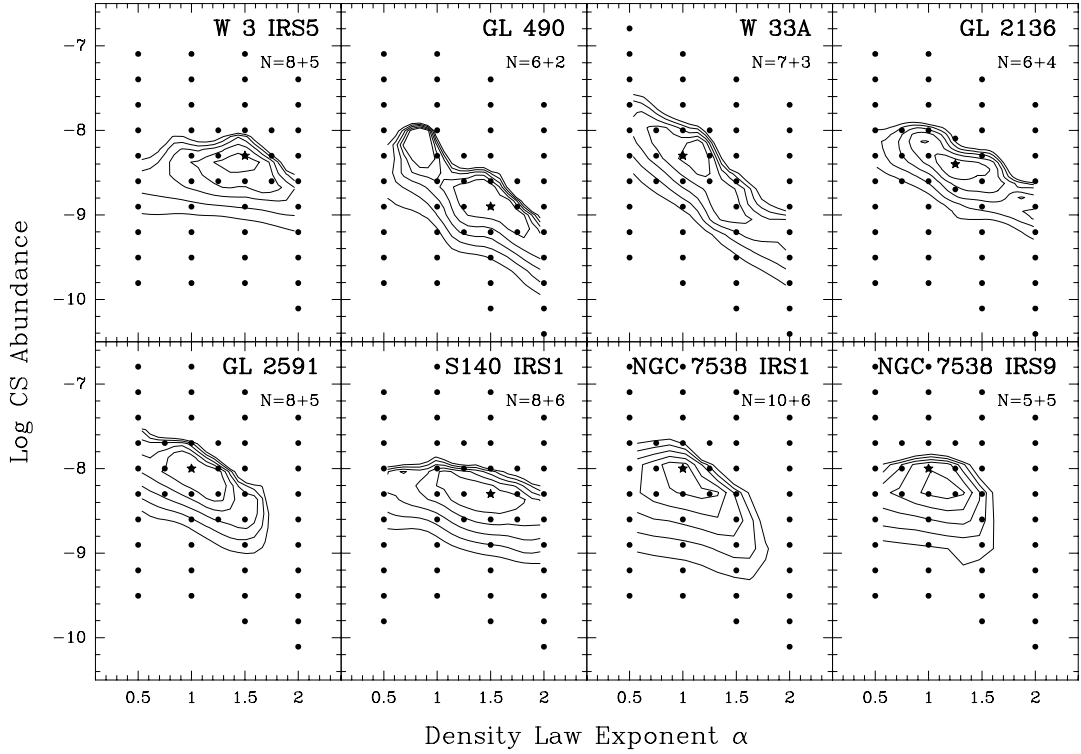,angle=-90,width=15cm}

    \caption{Fit quality parameter $\chi^2$ plotted as a function of
      density law exponent $\alpha$ and CS abundance. Contours
      increase by~1 and start at~1 for W~3~IRS5, GL~7009S, IRAS~20126,
      W~3~(\water) and NGC~6334~IRS1, 3 for NGC~7538~IRS1 and
      NGC~7538~IRS9, and at~2 for the other seven sources.
      The best-fitting model is indicated by a star.  In the top right
      corner, the number of CS and \cs\ lines included in the $\chi^2$
      calculation is listed.  }

    \label{f:chi}
\end{figure}

The results of the emission line models are presented in
Figure~\ref{f:chi}. The fit quality parameter $\chi^2$ is plotted as a
function of the density law exponent $\alpha$ and the CS abundance
with respect to \htwo.  The elongation of the $\chi^2$ contours
indicates that the quality of the fit depends to first order on the CS
abundance and to second order on the density profile. The parameters
of the best fitting models are summarized in Table~\ref{t:m_gas}.

It is seen from Table~\ref{t:m_gas} that for the main sample, $\alpha
= 1.0-1.5$, while the other sources have $\alpha = 1.75 - 2.0$.  This
result is consistent with the result from the dust models
(\S~\ref{s:mod_dust}), namely that for a given source, the model with
lowest $\alpha$ has the highest mid-infrared flux.  Thus, the
mid-infrared brightness of the main sample is not a pure orientation
effect.  However, since the spherical dust models fail to reproduce
the near-infrared emission of all sources, deviations from spherical
symmetry are important. This conclusion is supported by synthetic CS
line profiles from our power law models, which are self-absorbed for
all but the least massive sources. The effect of deviations from
spherical symmetry on line profiles and near-infrared emission was
discussed for GL~2591 in \citet{fvdt99}. Decreasing the central column
density by a factor of a few can increase the mid-infrared emission
and decrease the self-absorption substantially.

The best-fit CS abundance is $3\times 10^{-9}$ on average, with
source-to-source variations of a factor of~3 for most sources. The hot
cores and the ultracompact \ion{H}{2} region have the highest CS
abundances: $(1-2)\times 10^{-8}$. The low CS abundance in GL~7009S,
$0.4\times 10^{-9}$, is probably due to freezing out of the molecules
onto the dust grains, as suggested by the large column densities of
several ice species towards this source \citep{dhen96}.

Figure~\ref{f:chi} also lists the number of CS and \cs\ lines
observed.  The sources for which the most lines have been observed
also have the tightest constraints on $\alpha$. For less well-observed
sources, our results may be influenced by sampling effects. In
particular, if only a small range of molecular energy levels is
available, the value of $\alpha$ may be biased, or a gradient may be
hard to detect. We have checked for such effects by calculating
$\bar{J}$, the average upper J-level of the data set. Most sources
have $\bar{J}=4.5-5.5$, which implies that the observable range of
critical densities has been evenly sampled. However, for GL~7009S, and
IRAS~20126, for which fewer high-excitation line data exist,
$\bar{J}=3.2-3.6$, and the model results are therefore less robust.

The value $\alpha = 2$ found for the hot cores is higher than for the
other sources, and we have investigated if this result could be an
artifact of our assumption of a constant molecular abundance.  Hot
cores have high abundances of many molecules including saturated
organics, which are thought to be due to freshly evaporated grain
mantles.  In this picture, the radius where the temperature reaches
90~K is an important threshold because water ice, the main ingredient
of interstellar grain mantles, evaporates. It is possible that the
abundance of CS is enhanced above this temperature as well.  To test
this idea, a model has been run with the luminosity and the \mass\ of
W~3~(\water), but with $\alpha=1.5$. The CS abundance is $5\times
10^{-9}$ in the outer parts, but enhanced where the dust temperature
exceeds $90$~K. A match to the data of equal quality as with the best
constant-abundance model ($\alpha=2.0$, CS/H$_2=1\times 10^{-8}$:
Table~\ref{t:m_gas}) was obtained by enhancing the CS abundance in the
inner region by a factor of 10. This result indicates that the hot
cores may have a density structure similar to that of the main sample,
if plausible variations of CS/\htwo\ with radius are considered. SCUBA
maps of other ``hot core''-type sources by Hatchell et al.~(2000,
submitted to A\&A) indeed suggest values of $\alpha\approx 1.5$.

\subsection{Comparison of Dust and Gas Tracers}
\label{mixing}

\begin{figure}[hp]
  
  \psfig{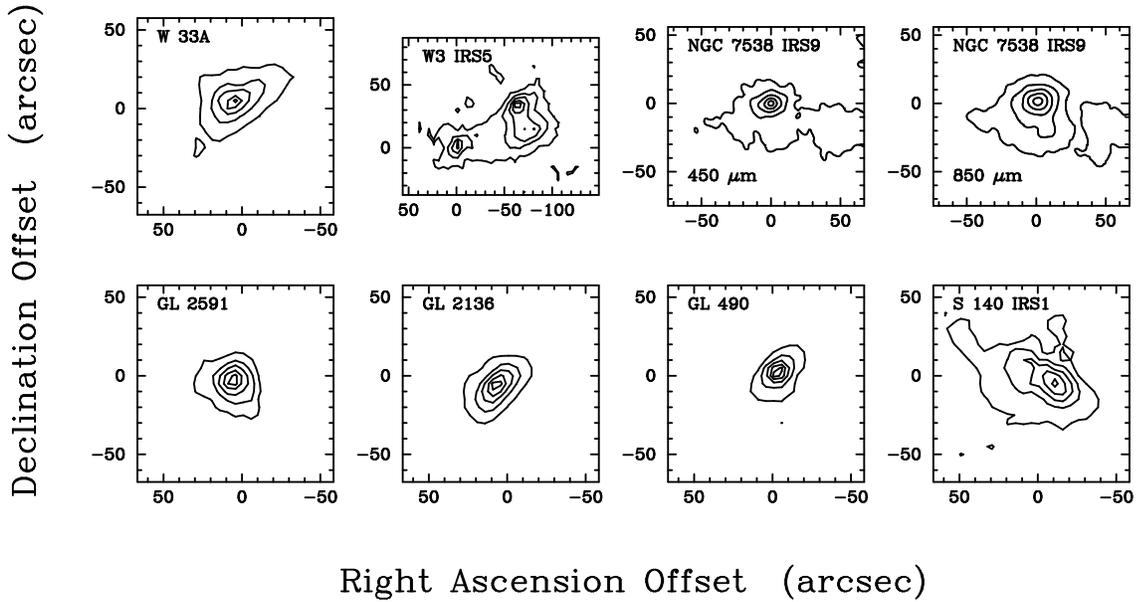}

  \caption{Maps of 350~\mic\ continuum emission made with SHARC, and 
   of NGC~7538 at 450 and 850~\mic, observed with SCUBA on the JCMT.
   Contours are  10 to 90\% of the peak brightness, in steps of
   10\%. Peak brightness (in Jy/beam) is 45.8 for W~33A, 7.7 for
   W~3~IRS5, 27.7 for NGC~7538~IRS9 at 450~\mic\ and 4.8 at 850~\mic,
   28.5 for GL~2591, 34.5 for GL~2136, 13.4 for GL~490 and 37.7 for
   S~140~IRS1.}
  \label{f:cmaps}
\end{figure}

\begin{figure}[hp] 
   \begin{center} 

   \psfig{figure=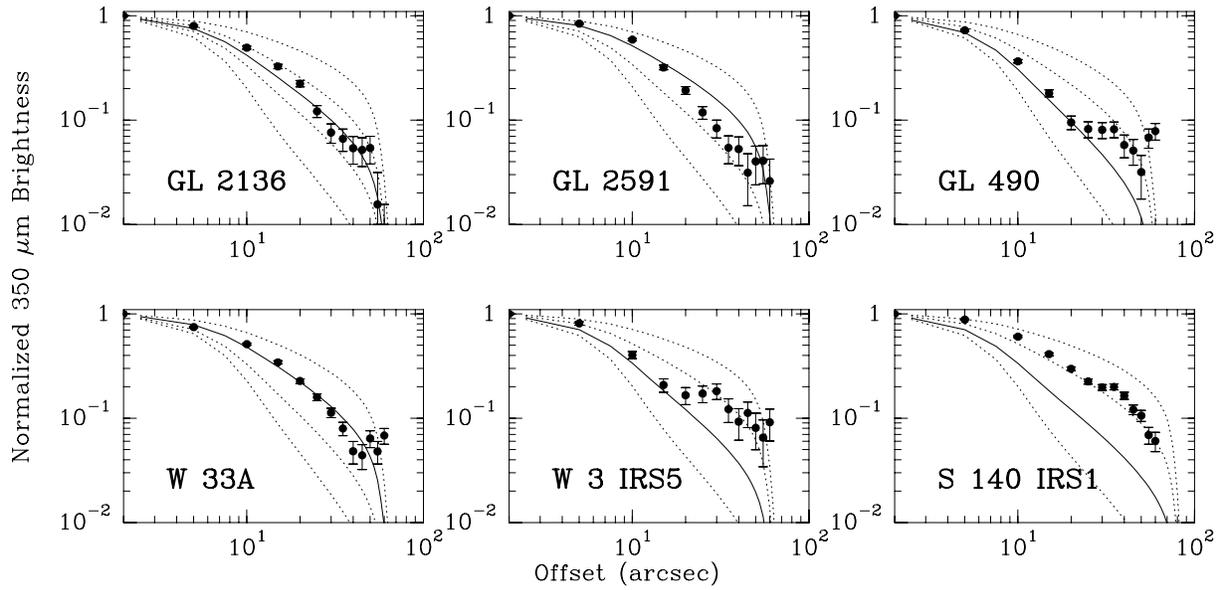,angle=-90,width=17cm}

\caption{Radial profiles of $350$~\mic\ emission observed with SHARC (a)
  and SCUBA (b). Dotted lines are model fits from the power law models for $\alpha=0.5,
  1.0, 1.5$ and $2.0$ ({\it top to bottom}). The solid line is the
  model that fits the CS excitation best.}

      \label{fig:dust_prof} 
      \end{center} 
   \end{figure}

The density structures derived from the CS excitation will now be
tested by using them to model the radial profiles of the submillimeter
dust emission. This emission is optically thin and hence measures the
column density distribution, given the calculated temperature
structure. The points in Figure~\ref{fig:dust_prof} are averages of
slices along the North-South and East-West directions through the
images shown in Figure~\ref{f:cmaps}. To reduce the noise, the data
have been folded about the image maximum. In the case of W~3~IRS5, the
local maximum in the SHARC map that corresponds to the infrared
position was used to center the slices, and the Western direction was
not included in the scans because of confusing extended emission
clearly visible in Figure~\ref{f:cmaps}.  Superposed are slices
through model images for various values of $\alpha$ as dotted lines,
while the solid line corresponds to the value of $\alpha$ derived in
\S~\ref{s:mod_gas} from the CS excitation. The model profiles (in a
$10''$ beam) were calculated with a code kindly supplied by L.~Mundy.
The models are exactly those used to produce Fig.~\ref{f:chi}, except
that for W~3~IRS5, GL~490 and NGC~7538~IRS9, the outer radii were
increased to $60''$ to avoid edge effects. In these cases, the radial
profile of the dust emission was quite extended.

Figure~\ref{fig:dust_prof} demonstrates that the dust and gas tracers
agree very well on the best value of $\alpha$, implying that the
volume density and column density distributions are consistent.  This
result indicates that the dust and gas are well-mixed and that the
structure of the envelopes is fairly homogeneous and not very clumpy.
The only exception to this rule is S~140~IRS1, for which the dust
gives $\alpha=1.0$ while the gas gives $\alpha=1.5$. Although an
inhomogeneous (clumpy) structure would cause a discrepancy in this
direction, it is hard to see why only one of our sources would have
such a structure.  Clumps in S~140 have been proposed by e.g.
\citet{spaan97}, but in a much larger ($\sim$arcmin-sized)
region outside the dense star-forming core.  More likely to be
important here is an elevated temperature caused by the nearby sources
IRS2 and IRS3 at $10-15''$ offsets, which have luminosities similar to
that of IRS1 \citep{evan89}, and by external heating by the nearby
ionization front \citep{lest86}.  So S~140 is unique in our sample in
that it is the only source where our assumption of one central heating
source breaks down.

Although the data on GL~2591 are well matched by the power law on
average, the slope of the data is shallower inside a radius of
$10-20''$, and drop more steeply outside this radius. This suggests a
variation in the temperature or the density gradient with
radius. Since the effect is more pronounced towards shorter
wavelengths, a variation of the temperature gradient with radius seems
more plausible. An inner region of roughly constant temperature, such as
caused by a central cavity with little or no extinction, may reproduce
the data, but such models fall outside the scope of this
paper.

\section{Discussion}
\label{s:disc}

We have modeled single-dish continuum and line data with power law
models $n=n_0 (r/r_0)^{-\alpha}$, and obtained good solutions ($\chi^2
= 1-3$) with $\alpha \approx 1.0-1.5$ for the bright mid-infrared
sources and $\alpha=1.5-2.0$ for the other sources, although the
highest value $\alpha=2.0$ found for two ``hot cores'' may be an
overestimate due to an enhanced CS abundance close to the center. This
section 
investigates the validity of the assumptions that went into the
models and discusses possible implications of their results.

\subsection{Envelope Structure}
\label{sec:struc}

\begin{figure}[hp]
  
   \psfig{figure=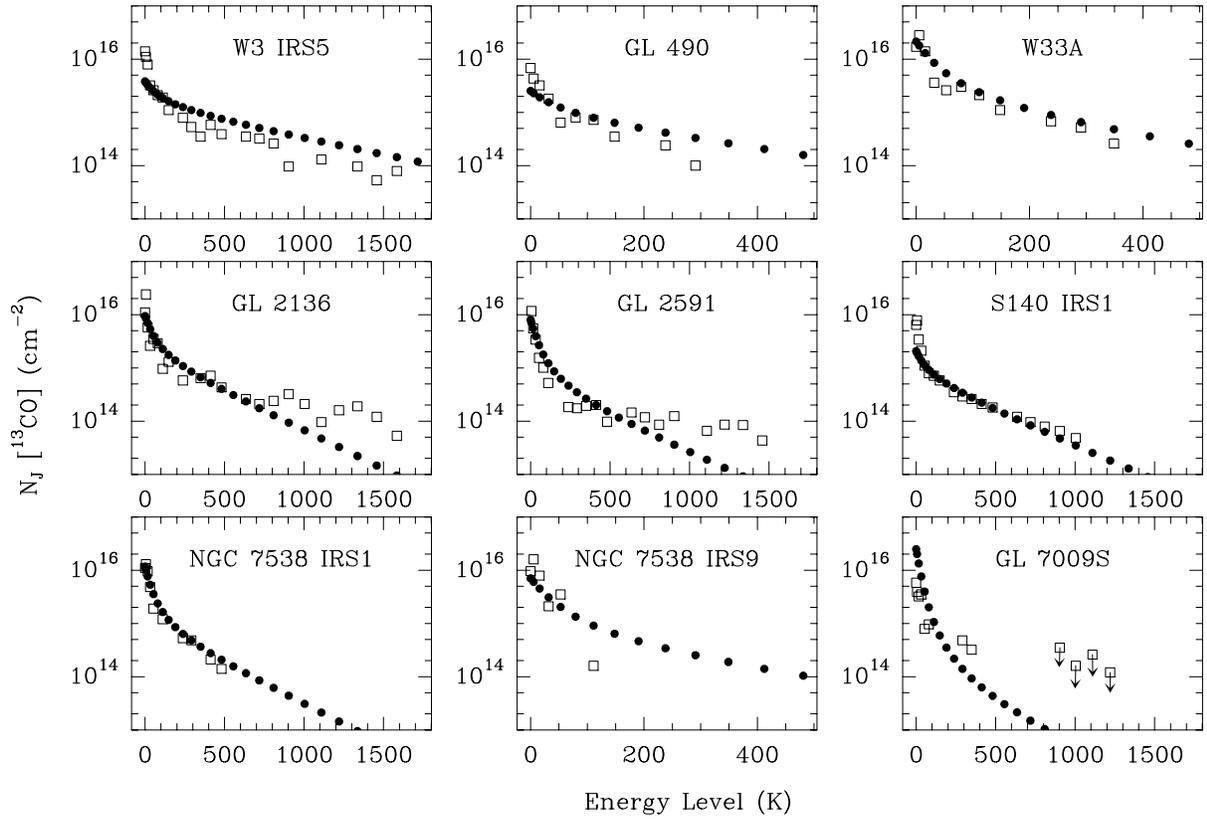,angle=-90,width=17cm}

  \caption{Observed and modeled column density in the rotational
    states of \co\ up to $J=25$ for the NIR-bright sources.  Data are
    from Mitchell et al.~(1990) except GL~7009S where they are from
    Dartois et al.~(1998). The model values have been scaled
    to the observed total column density. }  

   \label{fig:ir_abs} 
   \end{figure}

The models developed in Section~\ref{s:models} assume that the clouds
are spherical and homogeneous. \citet{mitc90} observed the \co\ 
$v=1\gets 0$ band at $4.7$~\mic\ in absorption towards the bright
mid-infrared sources. The excitation of the quiescent (non-outflowing)
gas rules out single-temperature models, but the data are well fitted
by a model with two discrete temperatures. The column density ratio of
these ``cold'' and ``hot'' components varies from~$\approx 1$ to
$\approx 5$. The physical origin of such a two-temperature structure
is not clear, however. We have calculated the column densities in each
rotational state of \co\ up to $J=25$ from our power-law models, and
the results are compared to the data of Mitchell et al.\ in
Figure~\ref{fig:ir_abs}. The low-resolution data for GL~7009S from
\citet{dart98a} may contain a contribution from outflowing gas, and
are hence regarded as upper limits.  The models have been scaled to
the observed total column density, which is always within a factor
of~2 from the column density measured by our submillimeter emission
data (\S~\ref{s:est_phys}). The good match (maximum deviation of a
factor of $3$) to the infrared data, which sample a pencil beam toward
the infrared source, indicates that a spherically symmetric model is a
reasonable first-order description of these sources. Together with the
evidence from the CS emission line profiles and the near-infrared
continuum emission, we estimate the deviations from spherical geometry
as a decrease in optical depth by a factor of $\approx 3$ over an
$\ltsim 10''$ area.

\begin{figure}[hp] 
   \begin{center} 

   \psfig{figure=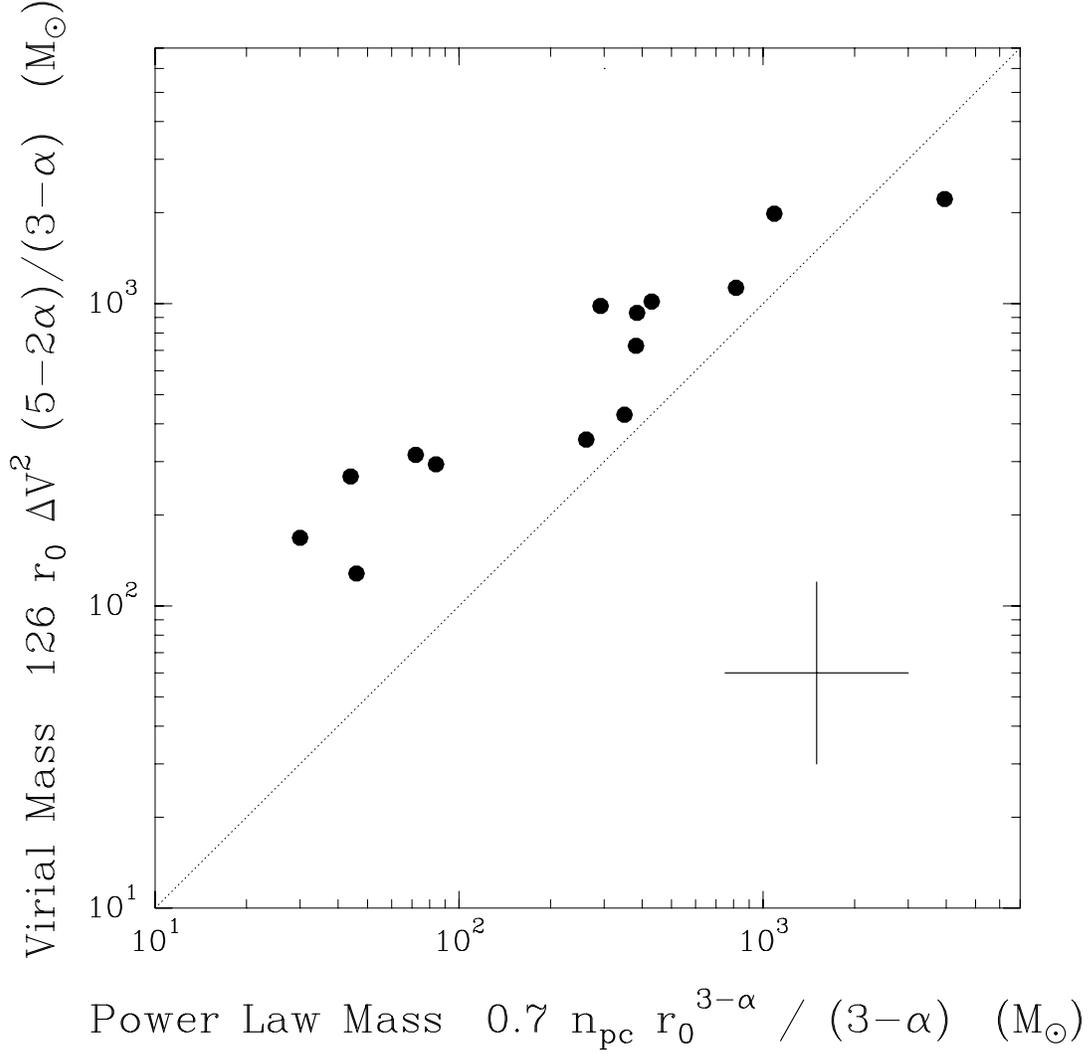,width=15cm}

\caption{Comparison of cloud masses
  derived from the power law models with those derived from the virial
  theorem. The quantity $n_{pc}$ is the density at a distance of 1~pc
  from the center; $\Delta V$ is the line width from
  Table~\ref{vlsr.tab} and the other quantities are in Table~\ref{t:m_gas}.}

      \label{fig:mvir} 
      \end{center} 
   \end{figure}

The assumption of a homogeneous density structure is tested in
Figure~\ref{fig:mvir}, where \mass\ is compared to the mass derived
from the virial theorem inside the same radius. The latter method
assumes that the cloud's gravitation just balances its pressure as
measured by the line width $\Delta V$ (Table~\ref{vlsr.tab}). If the
gas is clumpy, the virial mass is smaller than the power law mass by a
fraction $f_v$, with the volume filling factor $0<f_v<1$.  The ratio
$M_V/$\mass\ has a mean value of 2.77 and a standard deviation of
1.66. The only source for which \mass$>M_V$ is GL~7009S, for which
the model results are uncertain.

While there is no evidence for clumping within the envelopes, other
authors have found that regions of massive star formation are usually
clumped on a much larger scale. It may not be a coincidence that the
sizes derived for the clumps in S~140 by \citet{plum94} and in M~17~SW
by \citet{wang93}, $0.2$~pc, are similar to the sizes of the envelopes
found here. It is however outside the scope of this paper to
investigate a possible evolutionary connection between molecular cloud
clumps and the envelopes of massive young stars.

\subsection{Relation of CO Abundance with Temperature}
\label{s:ice}

Our assumed values of the gas to dust mass ratio and the dust opacity
at submillimeter wavelengths were tested by modeling C$^{17}$O $J=3\to
2$ and $2\to 1$ observations. The best-fit values for the abundance of
CO are listed in column~6 of Table~\ref{t:m_gas}. The derived
abundances are $10-60$\% of the value of $2.7\times 10^{-4}$
measured by \citet{lacy94} towards the warm cloud NGC~2024.  Our
CO/\htwo\ value for GL~2591 is in good agreement with the value of
$3.1\times 10^{-4}$ measured by C.~Kulesa (1999, priv.\ comm.), while
in the case of GL~490, our value is a factor of~7 lower.  The CO
abundance measured in emission is expected to be lower than in
absorption, because the emission data are more sensitive to extended
cold gas. When measured in infrared absorption, which samples warm
material close to the star, the column density ratio of solid to
gaseous CO never exceeds unity towards massive young stars
\citep{mitc90,evd98b}.

\begin{figure}[hp]
  \begin{center}
    
   \psfig{figure=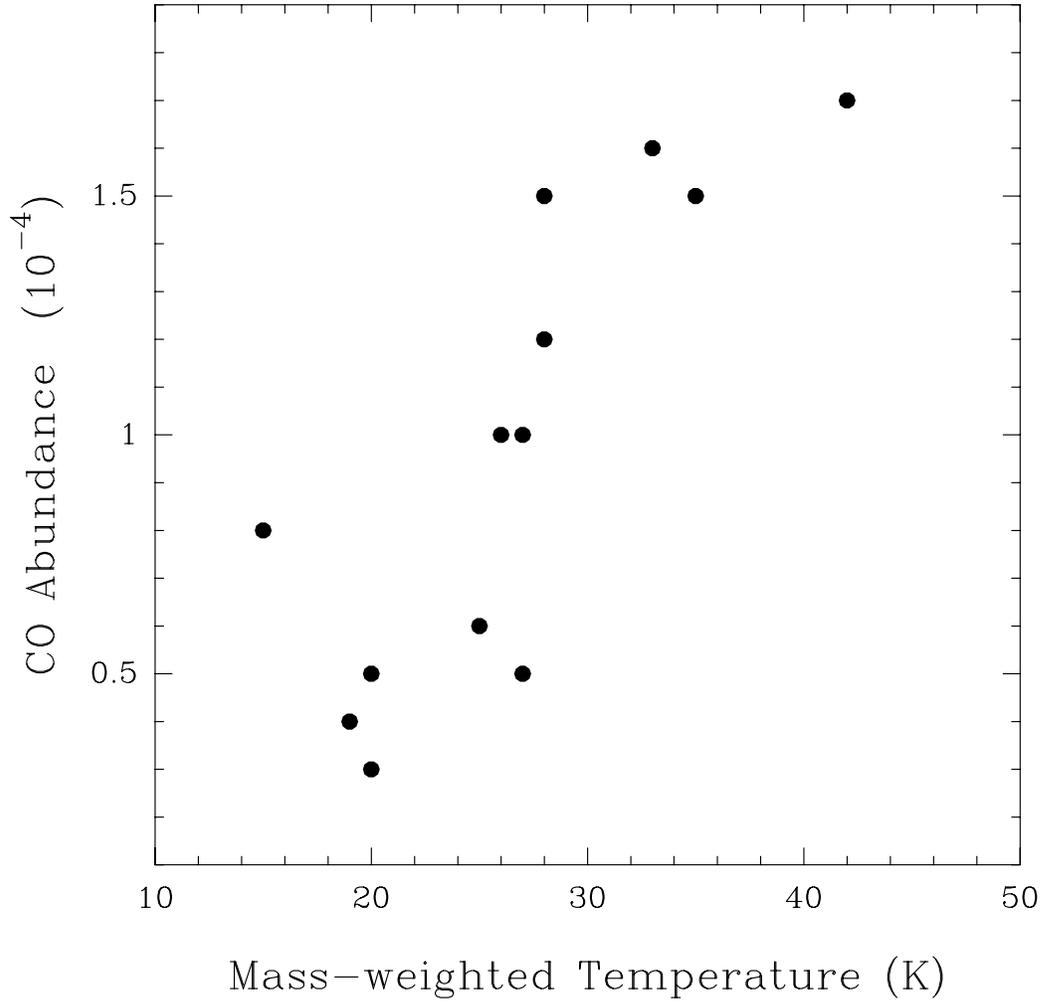,width=15cm}

    \caption{Abundance of CO derived from \co\ observations plotted against the mass-weighted
      temperature in our models.}
    \label{fig:coh2}
  \end{center}
\end{figure}

Although source-to-source variations in grain opacity cannot be ruled
out, a more likely explanation of the spread in CO abundances is that
different amounts of CO are frozen out on grain surfaces. This occurs
at temperatures $\ltsim 20$~K in the case of pure CO \citep{sand93},
and up to $\sim 45$~K for CO-\water\ mixtures. In both cases, a
correlation of the abundance with temperature is expected.
Figure~\ref{fig:coh2} shows a plot of the CO abundance derived from
C$^{17}$O observations versus the mass-weighted temperature, defined
as $ \bar{T} = {\int T(r) n(r) r^3 dr}/{\int n(r) r^3 dr}$ (see
Table~\ref{t:m_gas}).  The two quantities are seen to be correlated,
which we interpret as the effect of depletion of CO at low
temperatures.

These results strongly suggest that depletion and thermal desorption
are the processes controlling the abundance of CO in the gas phase,
after chemical reactions lock up almost all of the available carbon.
This situation is in contrast with that of CS, for which the abundance
shows no clear trend with $\bar{T}$. This difference presumably
reflects the much greater chemical inertness of CO compared to CS, and
the lower evaporation temperature of CO compared with CS.

In the case of GL~7009S, the \co\ column density observed in
absorption by \citet{dart98a} is similar to that in our model. This
implies that along this line of sight, CO is much less depleted than
CS, for which a depletion by a factor $\approx 10$ was found compared
to the other sources in our sample (Section~\ref{s:mod_gas}). Again,
this can be understood from the difference in evaporation temperature
for these two molecules.  The large column of CO ice towards, e.g.,
W~3~IRS5, is not predicted by our model. We suggest that this
star-forming core is surrounded by an extended dense cloud which also
produces the self-reversed CO emission profiles. Sources like GL~2591
and GL~2136 do not seem to have such a ``skin'', or they have at most
a translucent one.

\citet{menn98} found that the submillimeter opacity of dust grains
increases by $10-50$\% when the temperature rises from 24~to 300~K.
This effect would decrease the masses of our sources and increase the
inferred abundances of CO, and qualitatively produce the same trend as
seen in Figure~\ref{fig:coh2}. However, over the applicable
temperature range, $20-50$~K, an effect of only $5-25$\% should occur,
which is much less than the factor of~3 increase in CO abundance that
we find. Although the grain opacity is likely to vary within our
sources, they do not influence the conclusion that the CO abundance is
controlled by freezing and sublimation.

\subsection{The Inner $\ltsim 1000$~AU: Evidence for Compact
  Dense Material}
\label{sec:small}

In this section, the envelope models derived in \S~\ref{s:models} will
be compared to the OVRO continuum data. Since the images presented in
Figure~\ref{fig:ovromaps} show only compact, circularly symmetric
structure, a simple one-dimensional analysis in the Fourier plane is
sufficient. 

\begin{figure}[hp] 
   \begin{center} 
   \psfig{figure=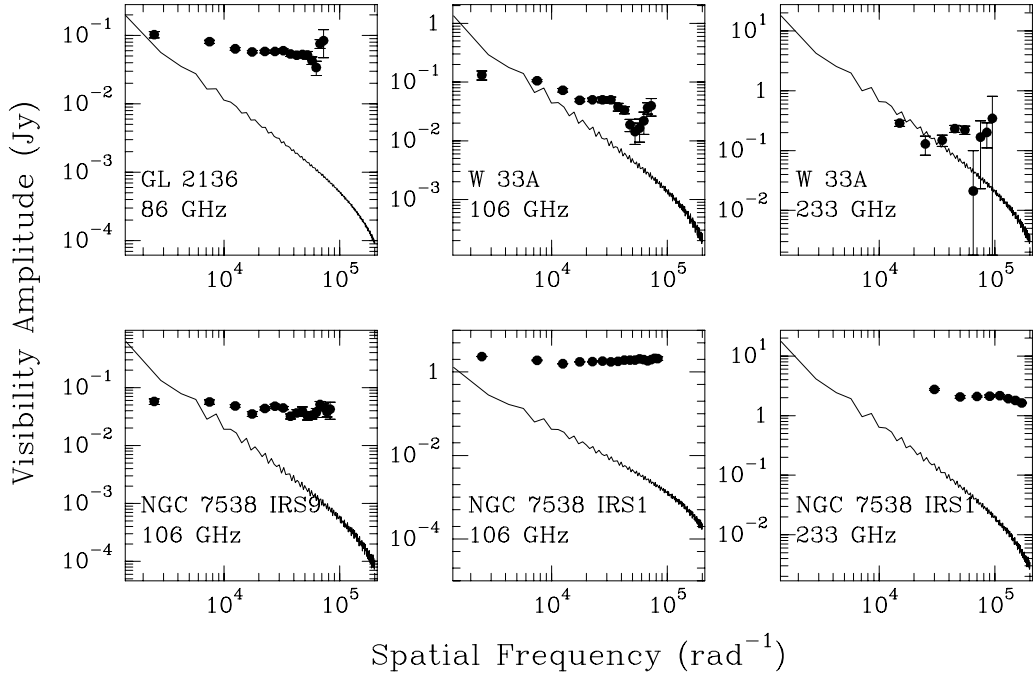,angle=-90,width=20cm}

\caption{Continuum visibilities observed with OVRO compared to the
  power law models.}
      \label{fig:cvis} 
      \end{center} 
   \end{figure}

The points in Figure~\ref{fig:cvis} are the OVRO data, averaged in
annuli around the source in the \textit{uv} plane. In the case of
W~33A, the source MM2 was subtracted from the data before averaging. 
The error bars represent the $1\sigma$ spread between the data points
in each bin, and do not include the overall calibration error. 
Superposed on the data points are model curves, derived by calculating
the dust emission from the best-fit models (Table~\ref{t:m_gas}) and
Fourier transforming the result.

Only at the lowest spatial frequencies, the data match the model
curves within the calibration error. Most data points lie well above
the model curves, and the observed amplitudes do not drop with
increasing spatial frequency, as the models do. These differences
suggest that the emission detected with OVRO is not related to the
envelopes seen in the single-dish data, but due to compact structure
of size $\ltsim 2''$. The same conclusion was reached for GL~2591 by
\citet{fvdt99}.

Towards the highest observed spatial frequencies, $0.5-1 \times
10^5$~rad$^{-1}$, corresponding to $\sim3000$~AU at 1~kpc, the compact
emission is $10-100$ times stronger than that of the envelope model.
This factor is a lower limit to the column density contrast between
the two components, since the compact emission is probably optically
thick, as indicated by the spectral indices. Combined with the
envelope column densities in Table~\ref{vlsr.tab}, this result
suggests that the compact emission seen with OVRO can explain the
asymmetry observed in the CS line profiles (\S~\ref{s:res:spectra}). 

Further constraints on the nature of the compact emission can be
obtained from the flux densities (Table~\ref{posflux.tab}), which
imply brightness temperatures of $50-80$~K for NGC~7538~IRS1 and
$\approx 1$~K for the other sources. These values imply that the
emission is beam diluted by a factor of $\gtsim 100$, since the
physical temperature of the compact structure must be at least that of
the surrounding envelope, which acts as an oven (see
\citealt{fvdt99}). Assuming a temperature of $200$~K for the compact
dust, a lower limit to the mass of $\sim 10$~\msol\ is obtained.  In
the case of NGC~7538~IRS1, where free-free and dust emission both
contribute at these frequencies \citep{wood89,akab92}, the physical
temperature may be $>100$~K, and the beam dilution correspondingly
more.  For the other sources, the compact emission is most likely due
to dust, perhaps in the form of a dense shell or of a circumstellar
disk. The implied radius of the emitting region, $\ltsim 300-600$~AU
at $2-4$~kpc, seems small to hide an entire outflow lobe, but
presumably, the dense part of the flow which emits in high-$J$ CS
lines is confined to its center.

\subsection{Comparison with Low-mass Objects}
\label{sec:low}

\begin{figure}[hp] 
    \begin{center} 
   \psfig{figure=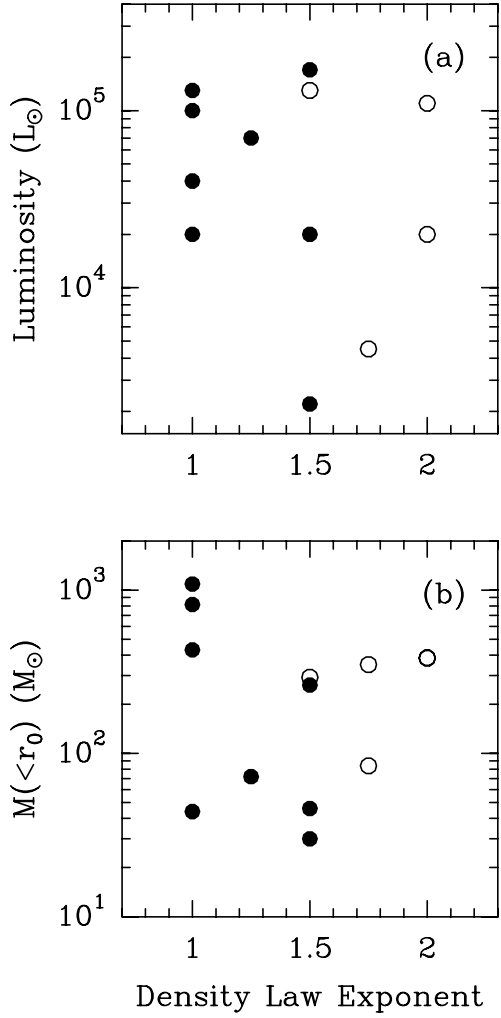,angle=-90,height=15cm}

 \caption{Slope of density gradient $\alpha$ versus source luminosity
   and mass. Filled symbols indicate the main sample,
   open symbols the comparison sample. The plotted quantities are
   explained in the text.}
       \label{fig:alpha} 
       \end{center} 
    \end{figure}

This section discusses possible origins of the spread in $\alpha$
within our sample.  First, the value of $\alpha$ may be related to
other physical parameters, in particular luminosity or envelope mass.
The structure of the material surrounding pre-main sequence stars of
low ($\sim$~solar) and intermediate ($\sim 3-8$~solar) mass has
received considerable attention in the recent literature: e.g.
\citet{ladd91,butn91,natt93,butn94,hoger99}.  These studies generally
find $\alpha \approx 2 \pm 0.3$ inside radii of 0.1~pc, masses of
$\leq$10~\msol\ and mean densities of $10^4$~\ccm. Of these
properties, only the radii are similar to those of the objects studied
here; the masses and mean densities are at least two orders of
magnitude smaller. The density gradients are significantly steeper
than found in this paper, but the gradients for intermediate-mass
stars are also lower than those for low-mass stars
\citep{natt93}. Indeed, in an earlier study of regions of high-mass
star formation, \citet{zhou94} suggested that more massive
star-forming regions tend to have flatter density distributions.
However, Figs.~\ref{fig:alpha}a and~b show that within our sample,
$\alpha$ does not appear correlated with source luminosity or envelope
mass. We did not find a relation with any other physical parameters
either, such as turbulent pressure as measured by the line width. 

Power laws have been proposed for the density structures of the
envelopes of young stars, with the index depending on the dominant
term in the pressure. If support against gravitational collapse is
primarily due to thermal pressure, a value of $\alpha=2$ is expected,
while if the dominant pressure is of nonthermal origin, the density
structure should follow an $\alpha=1$ law. In between these two cases,
a continuum of solutions can be constructed
\citep{lizan89,myer92,mclaug97}. Our observations may thus indicate
that the cores where massive stars form differ from those where
low-mass stars form in that they are supported against collapse by a
different mechanism. Within this interpretation, the spread in
$\alpha$ may be due to source evolution from an ($\alpha=1.0$) logatrope
to a collapsing region ($\alpha=1.5$).

\subsection{Tracing Envelope Evolution}
\label{sec:evol}

Despite having similar radio and infrared properties our sources show
different degrees of dispersal and warming-up of their envelopes.
This is shown by \citet{boog00} and \citet{evd98b} for a source sample
similar to ours based on several tracers: the far-infrared color, the
fraction of crystalline CO$_2$ ice, the gas/solid ratios of CO, CO$_2$
and H$_2$O, and the excitation temperature of CO. These quantities all
trace the envelope temperature, but all in a different way: some trace
the dust, others the gas, and some trace small scales and others large
scales. As shown by Boogert et al., the temperature indicators
correlate with each other.  Of particular interest is the fraction of
crystalline CO$_2$ ice, because crystallization is an irreversible
process, so that this indicator measures the maximum temperature,
while the others measure the current temperature.  Hence, the
temperature variations are not random fluctuations (like in FU~Orionis
objects), but reflect a systematic increase with time of the
temperature of the envelope as a whole.  

\begin{figure}[hp] 
    \begin{center} 

   \psfig{figure=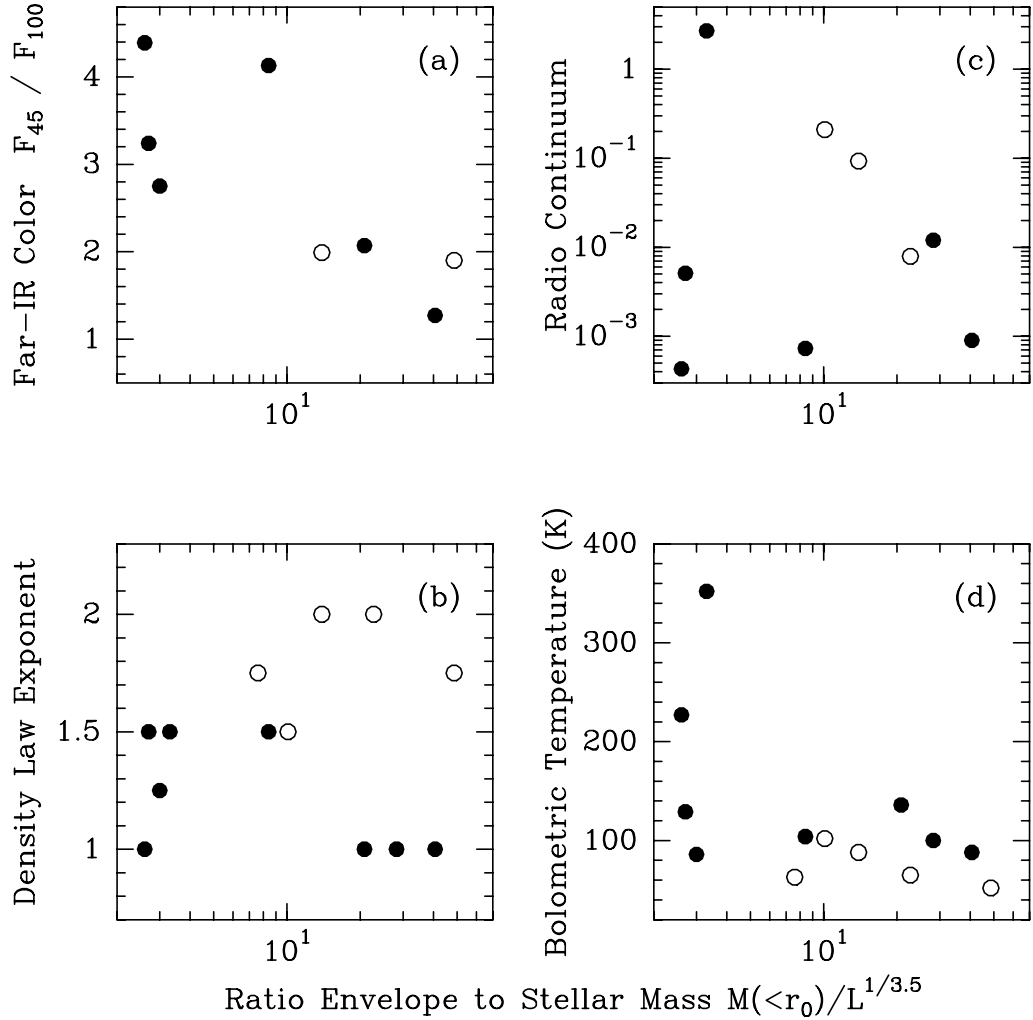,angle=-90,height=15cm}

 \caption{Possible evolutionary indicators plotted versus ratio of
   envelope mass to stellar mass: far-infrared color \textit{(a)},
   slope of density gradient $\alpha$ \textit{(b)}, normalized radio
   continuum flux density \textit{(c)} and bolometric temperature
   \textit{(d)}. Filled symbols indicate the main sample, open symbols
   the comparison sample. See the text for details.}
       \label{fig:evol} 
       \end{center} 
    \end{figure}

Figure~\ref{fig:evol}a plots one of these indicators, the far-infrared
color, versus the ratio of \mass\ to stellar mass, measured as
$L^{1/3.5}$. This quantity is equivalent to the ratio of submillimeter
to bolometric luminosity, often used to trace the evolution of young
low-mass objects \citep{andr99}.  The color $F_{45}/F_{100}$ is the
ratio of the flux densities at $45$~and $100$~\mic, both observed with
ISO-LWS in an $80''$~beam, which is large enough to cover the entire
envelope. These data were provided by A.~Boogert (1999, priv.\ comm.).
The plotted quantities are seen to be correlated, with higher
temperatures (traced by a larger $F_{45}/F_{100}$) corresponding to
lower values of \mass$/L^{1/3.5}$. Thus, the temperature variations
are indeed due to dispersal of the circumstellar material and can be
used to trace evolution.

Interestingly, $\alpha$ is not seen to be correlated with far-infrared
color (Fig.~\ref{fig:evol}b), and we conclude that on the time scale
of the dispersal of the envelopes of massive young stars, the density
structure of the envelopes does not change significantly. The masses
of the compact sources detected by us and by \citet{wood89} and
\citet{wyrow99b} do not show a relation with \mass/$L^{1/3.5}$ either,
although the uncertainty in these masses is rather high.

In addition to \mass$/L^{1/3.5}$, we have considered two
``derivative'' evolutionary indicators. The first is the bolometric
temperature \tbol, introduced for low-mass objects by \citet{myer93}.
As the envelopes in those systems are dispersed (probably by the
bipolar outflows), their bolometric temperatures rise because the
far-infrared emission decreases while the near-infrared emission
increases \citep{myer98}.  This process makes \tbol\ increase
monotonically from $\sim 30$~K to $3000$~K on a time scale of
$10^6$~yr. For all but two of our objects, values of \tbol\ based on
the spectral energy distributions of Fig.~\ref{f:dusto}, are
$50-150$~K.  This small spread suggests that all the sources are at
similar evolutionary stages; \tbol\ may also work less well than
$F_{45}/F_{100}$ because near- and mid-infrared emission have a
dependence on source orientation.

Another potential evolutionary indicator is the radio continuum flux,
which for a given luminosity and distance is expected to increase with
time as the dusty envelope is cleared away and the Lyman continuum
photons can all go into ionizing the gas without being absorbed by
dust. We have normalized the radio continuum flux densities in
Table~\ref{sam.tab} to the value expected for an H~II region ionized
by a main sequence star of the same luminosity and at the same
distance as the source, following \citet{chur93b}. However,
Figs.~\ref{fig:evol}c and~d show that neither the radio continuum
emission nor the bolometric temperature correlates well with
\mass/$L^{1/3.5}$, suggesting these are less useful evolutionary
indicators for very young massive objects than 
$F_{45}/F_{100}$.  Possibly the radio emission arises in a wind
(\S~\ref{s:samp}), or the stellar surface temperatures are
significantly below main sequence values.

We conclude that dispersal of the envelopes underlies the evolution of
both high-mass and low-mass objects, but that the observational
appearance is very different.  The dispersal process can be understood
in more detail by recalling that the brightest mid-infrared sources in
our sample are the weakest at radio wavelengths and vice versa
(Table~\ref{sam.tab}). This anticorrelation suggests that the erosion
of the envelope proceeds from the inside out. Massive stars are
efficient at removing circumstellar dust by their ionizing UV
radiation and by their strong winds. When the innermost, hottest
region of dust disappears, the mid-infrared emission will decrease,
but the radio continuum will increase because a larger region can be
ionized. The compact dust component detected with OVRO is optically
thick in the bright mid-infrared sources but thin in the others, which
trend may also trace dispersal of hot dust.  The far-infrared emission
will not be affected since it arises on larger scales. These trends
are consistent with the properties of our sample, although other
processes may play a role as well.

\section{Summary and Conclusions}
\label{s:conc}

We have presented maps and spectra of 14 regions of massive star
formation in continuum and molecular lines at submillimeter
wavelengths. The data are used to develop models of the physical
structure of the circumstellar envelopes on $10^2-10^5$~AU scales. The
column density is derived from the submillimeter continuum flux
densities, and the temperature structure is calculated
self-consistently using the size of the CS emission and the sources'
luminosities and distances from the literature. The density structure
is constrained with emission lines of CS, \cs\ and \hhco. The
following main conclusions are found:
 
1. The physical structure of the envelopes of deeply embedded massive
   young stars is characterized by radii of $3-9 10^4$~AU, masses of
   $10^2-10^3$~\msol\ inside these radii and visual extinctions of $\sim
   10^2-10^3$~magnitudes.  Temperatures increase from $\sim 20$~K at
   the outer edge to several $100$~K at $\sim 10^2$~AU from the star;
   densities increase from $\sim 10^4$~\ccm\ to $\sim 10^7$~\ccm.  The
   slope of the density gradient $\alpha$ is $1.0-1.5$, significantly
   smaller than the value of $\approx 2$ usually found for low-mass
   objects. This difference in density structure may be due to
   nonthermal pressure resisting gravitational collapse, while thermal
   pressure dominates in lower-mass objects. An $\alpha=2$ density
   structure is found here for two hot cores, but this is an
   upper limit since the CS abundance may be enhanced in
   the warm gas close to the star. 

   2. The shapes of the envelopes deviate somewhat from spherical, as
   shown by modeling infrared absorption line data of \co, the
   near-infrared continuum and the CS line profiles. The data are
   consistent with a decrease of the optical depth by a factor of
   $\approx 3$ in the central $\ltsim 10''$ area. The brightness of
   our main sample at mid-infrared wavelengths is partly due to this
   optical depth effect, but also to the flatter density structure.
   Inhomogeneities on top of the power law structure are small, since
   the masses obtained by integrating the power law models agree with
   masses found from the virial theorem to a factor of~3. The values
   of $\alpha$ found from CS are verified by model fits to the maps of
   dust emission which trace the column density distribution. The only
   exception is S~140~IRS1, probably because our assumption of a
   single central heating source is invalid in this case.
   
   3. Modeling of C$^{17}$O emission lines shows that $\approx
   40-90$\% of the CO gas is depleted onto dust, assuming a dust
   opacity of $\approx 1$~\scm~g$^{-1}$ at $\lambda 1300$\mic, as
   found previously for one of our sources, GL~2591 \citep{fvdt99}.
   The derived CO abundances correlate well with the mass-weighted
   temperature in our models.  This result suggests that in these
   sources, freeze-out and thermal desorption control the gas-phase
   abundance of CO. The CS abundance is $3\times 10^{-9}$ on average,
   ranging from $0.4\times 10^{-9}$ in the cold source GL~7009S to
   $1-2\times 10^{-8}$ in the ``hot core''-type objects.
 
   4. Dense outflowing gas is seen in the CS and H$_2$CO line wings,
   but much more so at blue- than at redshifted velocities. This
   asymmetry may be due to absorption by dense, optically thick
   material within 10$''$ from the star. The opacity of the power law
   models is a factor of $10-100$ below that required for this
   absorption. However, interferometric continuum observations at
   $\lambda 1300-3500$\mic\ show compact emission, probably from an
   $0\farcs3$ diameter optically thick dust component. The emission is
   a factor of $\sim10-100$ stronger than expected for the envelopes
   seen in the single-dish data, and may be due to a dense shell or a
   circumstellar disk. This component may explain the asymmetric CS
   and \hhco\ line profiles.

   5.  The evolution of these high-mass sources is traced by the overall
   temperature (measured by the far-infrared color), which increases
   systematically with decreasing ratio of envelope mass to stellar
   mass. The observed anticorrelation of near-infrared and radio
   continuum emission suggests that the disruption of the envelope
   proceeds from the inside out. Conventional tracers of the evolution
   of low-mass objects do not change much over this narrow age range.
   We conclude that the evolution of high-mass and low-mass envelopes
   have the same underlying mechanism (envelope dispersal), despite
   their different observational properties.

\acknowledgements

The authors are grateful to the staffs of the OVRO, JCMT, CSO and
IRAM~30m telescopes, in particular Remo Tilanus and Fred Baas at the
JCMT, Gilles Niccolini at IRAM and Leonardo Testi, Christine Wilson
and Debra Shepherd at OVRO.  We are also indebted to Michiel
Hogerheijde for developing the Monte Carlo code, to G\"oran Sandell
for providing the SCUBA maps of NGC~7538 and for his constructive
referee report, to Lee Mundy for providing the code to model dust
emission profiles, to Yancy Shirley who reduced the SHARC data, to
Adwin Boogert for providing the ISO-LWS data and to Craig Kulesa for
sending his CO data. This work also benefitted from discussions with
Malcolm Walmsley, Ed Churchwell and Xander Tielens.

This research is supported by NWO grant 614.41.003.  GAB gratefully
acknowledges support provided by NASA (grants NAG5-2297, -4813). NJE
gratefully acknowledges support from the State of Texas and NASA grant
NAG5-7203. We also wish to thank the people who maintain the
bibliographic databases at CDS (Strasbourg) and ADS (Harvard), of
which we have made extensive use.

\pagebreak


\begin{figure}[hp]
   \psfig{figure=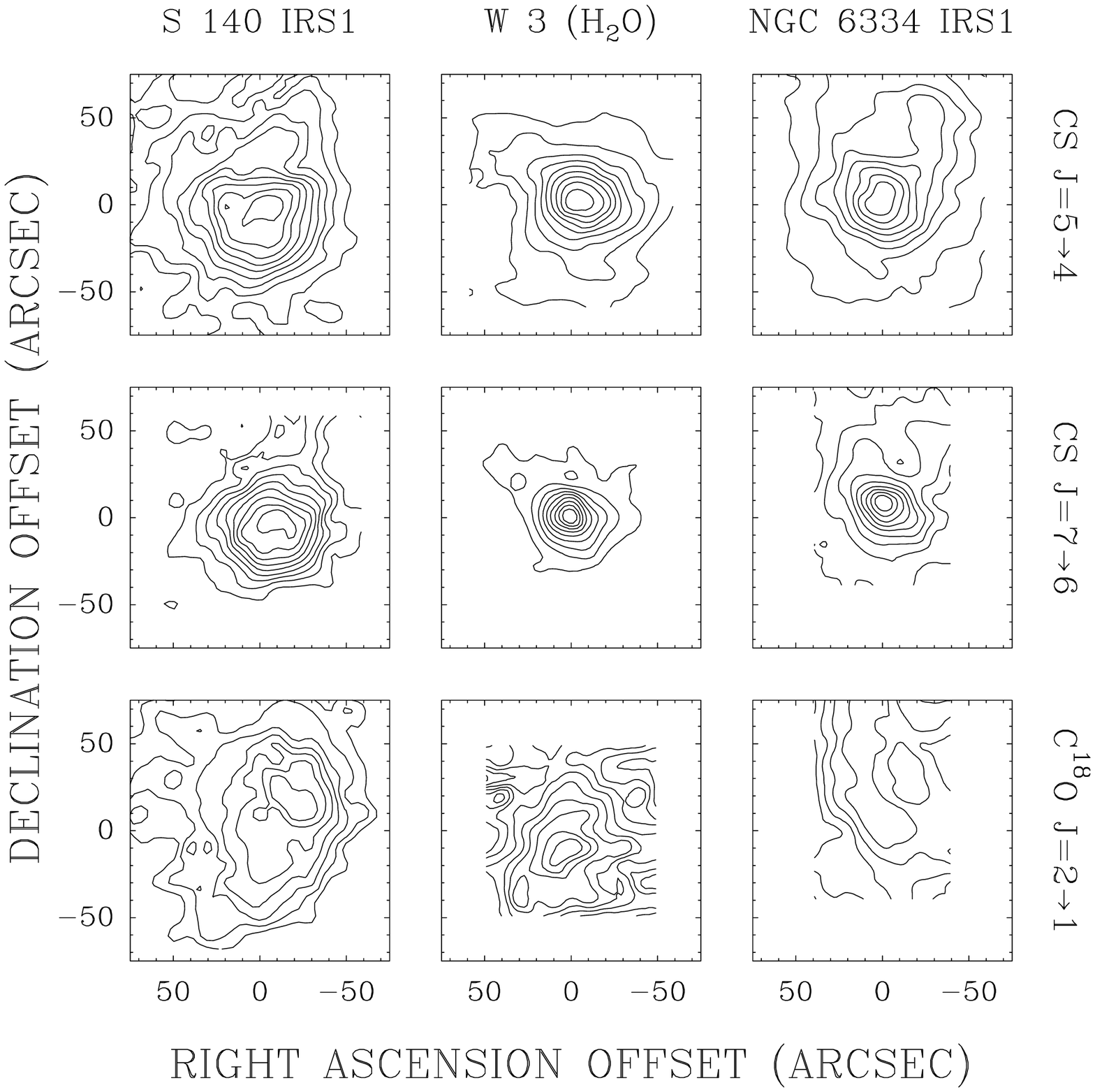,width=15cm}
   \centerline{\bf Figure 3 (continued)}
\end{figure}

\begin{figure}[hp]
   \psfig{figure=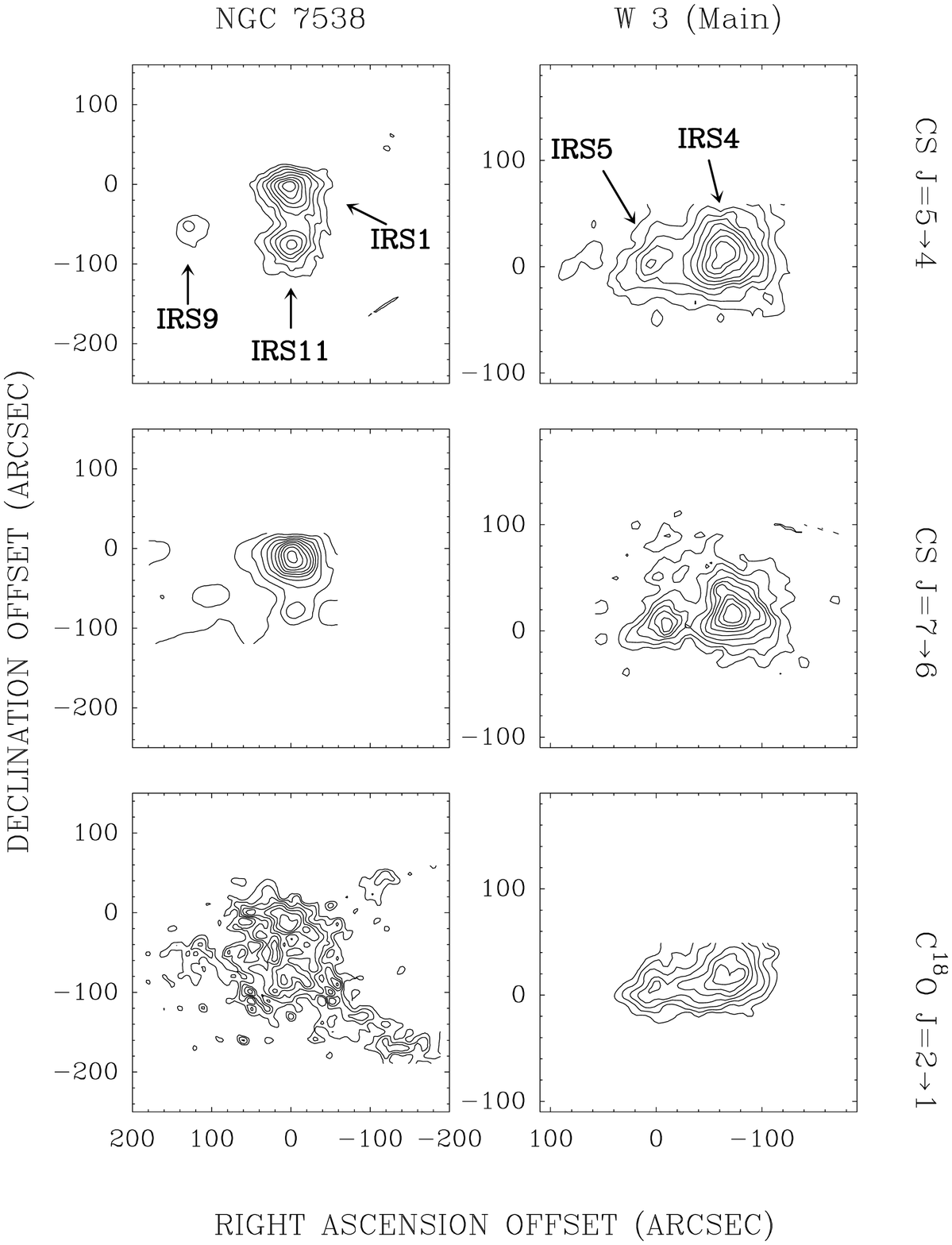,width=15cm}
   \centerline{\bf Figure 3 (continued)}
\end{figure}

\begin{figure}[hp]
   \psfig{figure=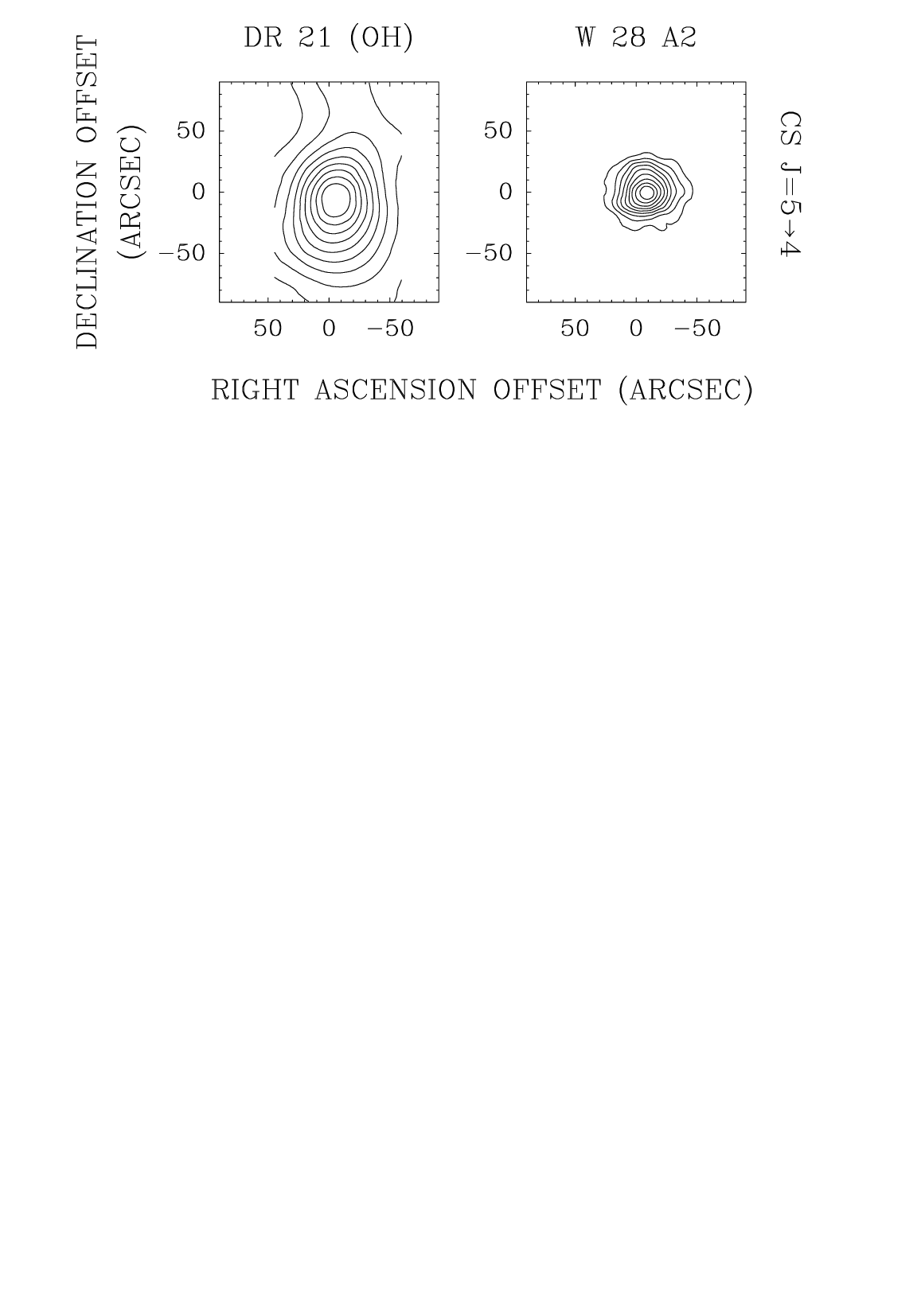,width=15cm}
   \centerline{\bf Figure 3 (concluded)}
\end{figure}

\pagebreak

\begin{figure}[hp]
   \psfig{figure=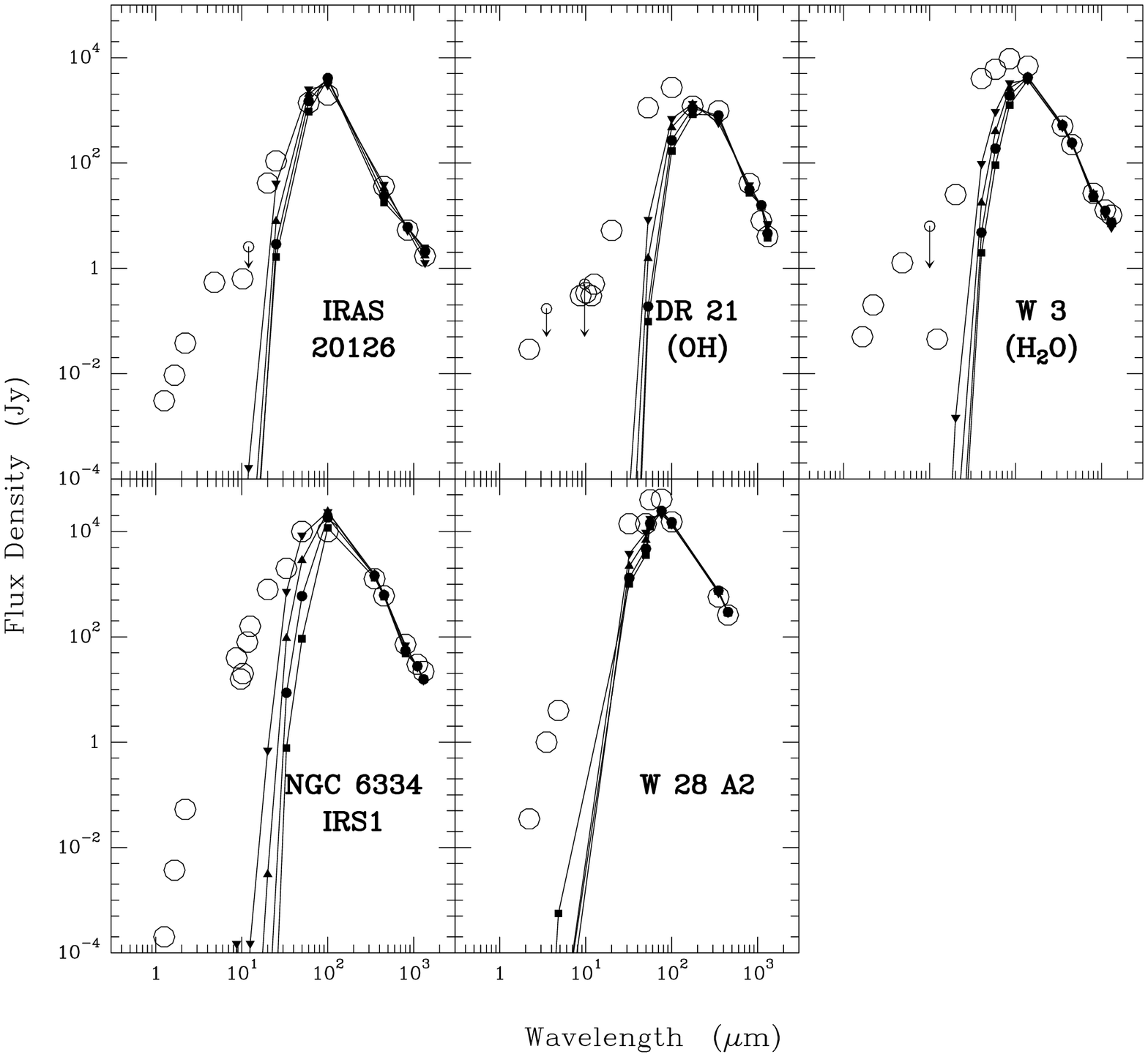,angle=-90,width=17cm}
   \centerline{\bf Figure 6 (continued)}
\end{figure}

\begin{figure}[hp]
   \psfig{figure=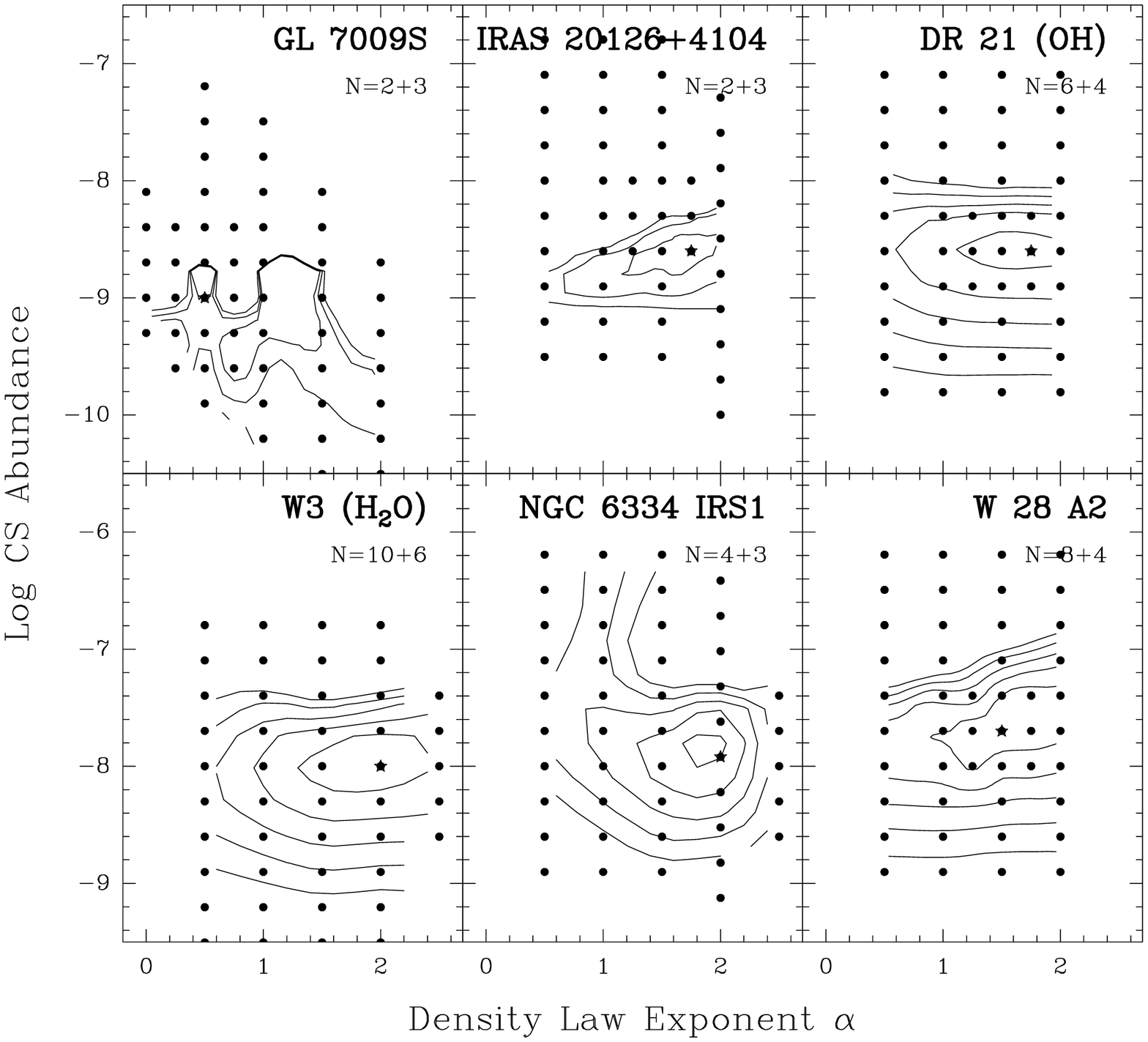,angle=-90,width=15cm}
   \centerline{\bf Figure 7 (continued)}
\end{figure}

\begin{figure}[hp]
   \psfig{figure=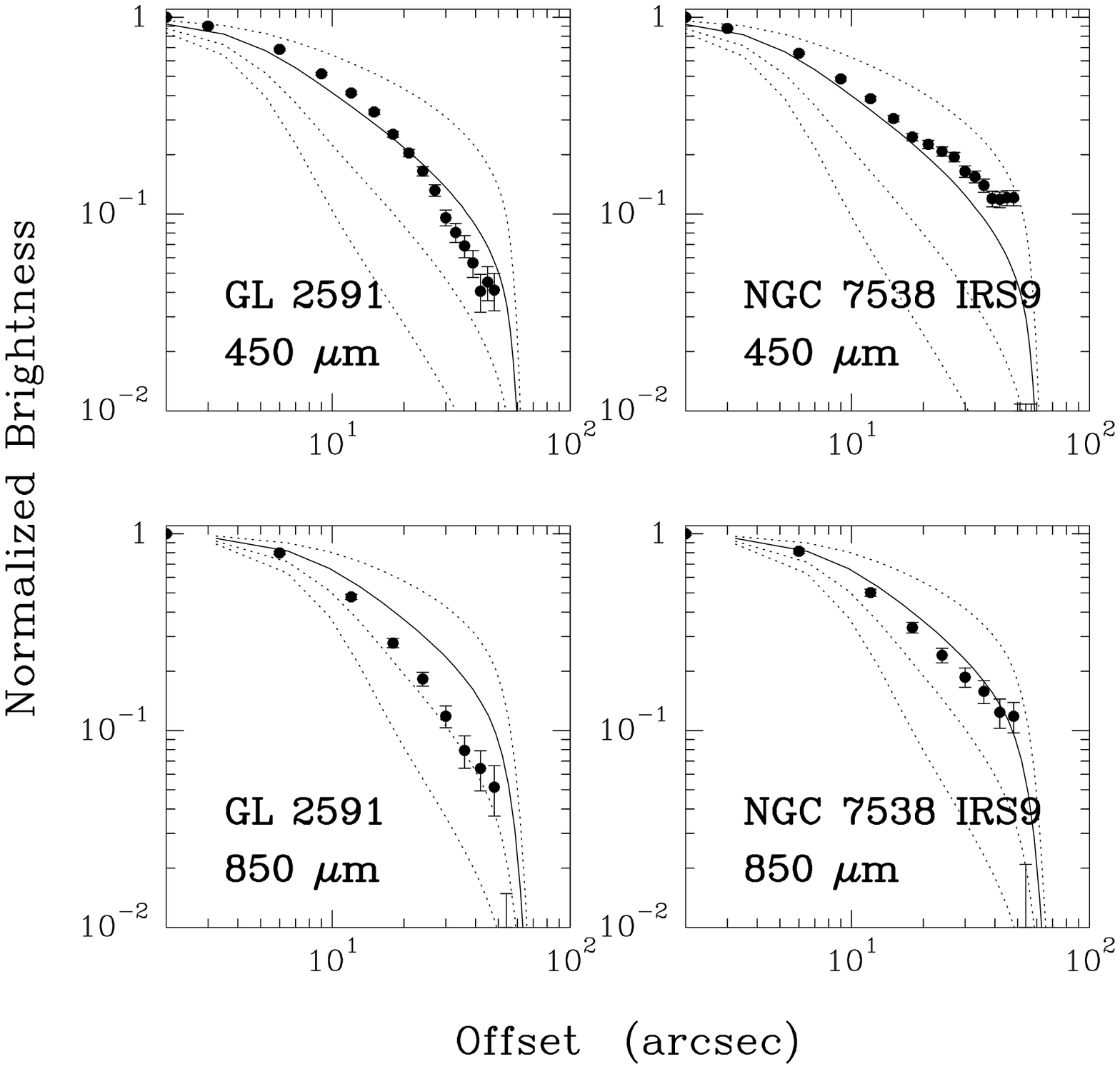,angle=-90,width=15cm}
   \centerline{\bf Figure 8 (continued)}
\end{figure}

\pagebreak


\begin{deluxetable}{llllllll}
\tablewidth{0pt}
\tabletypesize{\scriptsize}
\tablecaption{Source sample.\label{sam.tab}}

\tablehead{
\colhead{IRAS PSC} &
\colhead{Name} &
\colhead{RA (1950)} & 
\colhead{Dec (1950)}  & 
\colhead{Luminosity} & 
\colhead{Distance} & 
\colhead{$S_\nu$ (6~cm)} & 
\colhead{References} \\
 &
 & 
 &  
 & 
\colhead{(L$_\odot$)} & 
\colhead{(kpc)} & 
\colhead{(mJy)} &
       }
\startdata
 & & & & \\
\multicolumn{8}{c}{MIR-bright massive YSOs } \\
 & & & & \\
02219+6152  & W~3 IRS5         &$02^h$ $21^m$ $53\fs1$   &$+61^\circ$ $52'$ $20''$ &      $1.7\times 10^5$ & $2.2$ & 10.7 & 1,2,21 \\
03236+5836  & GL 490          &$03^h$ $23^m$ $38\fs9$   &$+58^\circ$ $36'$ $33''$ &       $2.2\times 10^3$ & $1$   & 3.2 & 3,4,22 \\
18117--1753 & W~33A            &$18^h$ $11^m$ $43\fs7$   &$-17^\circ$ $53'$ $02''$ &      $1\times 10^5$   & $4$   & 1.9 & 5,23\\
18196--1331 & GL 2136         &$18^h$ $19^m$ $36\fs6$   &$-13^\circ$ $31'$ $40''$ &       $7\times 10^4$   & $2$   & \nodata & 6 \\
18316--0602 & GL 7009S        &$18^h$ $31^m$ $41\fs6$   &$-06^\circ$ $02'$ $35''$ &       $3\times 10^4$   & $3$   & 2.7 & 7 \\ 
20275+4001  & GL 2591         &$20^h$ $27^m$ $35\fs8$   &$+40^\circ$ $01'$ $14''$ &       $2\times 10^4$   & $1$   & 0.4 & 8,9,24 \\ 
22176+6303  & S~140 IRS 1      &$22^h$ $17^m$ $41\fs08$  &$+63^\circ$ $03'$ $41\farcs6$ & $2\times 10^4$   & $0.9$ & 5.8 & 10,11,25 \\
23116+6111  & NGC 7538 IRS 1  &$23^h$ $11^m$ $36\fs7$   &$+61^\circ$ $11'$ $50\farcs8$ &  $1.3\times 10^5$ & $2.8$ & 111 & 12,13,26 \\
23118+6110  & NGC 7538 IRS 9  &$23^h$ $11^m$ $52\fs8$   &$+61^\circ$ $10'$ $59''$ &       $4.0\times 10^4$ & $2.8$ & $<0.5$& 12,13,23 \\
  & & & & \\
\multicolumn{8}{c}{MIR-weak massive YSOs} \\
 & & & & \\
20126+4104  & IRAS 20126+4104 &$20^h$ $12^m$ $41.0^s$   &$+41^\circ$ $04'$ $21''$      & $4.5\times 10^3$ & $1$ & $<0.3$ & 14,9,27 \\
\nodata     & DR 21 (OH)       &$20^h$ $37^m$ $14.2^s$   &$+42^\circ$ $12'$ $11''$      & $1\times 10^3$   & $1$ & $<10$ & 15,9,28 \\
02232+6138  & W~3 (\water)     &$02^h$ $23^m$ $17.3^s$   &$+61^\circ$ $38'$ $58''$ & $2 \times 10^4$ & $2.2$ & 1.5 & 16,2,29 \\
17175--3544 & NGC 6334 IRS 1  &$17^h$ $17^m$ $32.0^s$   &$-35^\circ$ $44'$ $05''$ & $1.1\times 10^5$ & $1.7$ & 1200 & 17,18,30 \\
17574--2403 & W~28 A2         &$17^h$ $57^m$ $26.8^s$   &$-24^\circ$ $03'$ $57''$ & $1.3\times 10^5$ & $2.0$ & 2100 & 19,20,31 \\
            & (=G5.89-0.39) & & & & & \\   
\enddata

\tablecomments{The first reference is for the luminosity, the second
  for the distance, and the third for the radio data.
  Spectrophotometric distances are given to 1 decimal place, and the
  corresponding luminosities are accurate to a factor of~2. The other
  distances are kinematic, in which case the luminosity is uncertain
  by a factor of~4. The kinematic distances to W~33A and GL~2136 were
  derived from data presented in this paper.}

\tablerefs{(1) \citealt{ladd93}; (2) \citealt{hump78}; (3) \citealt{chin91}; (4) \citealt{harv79}, but see
\citealt{snel84}; (5) \citealt{guer91}; (6) \citealt{kast94}; (7) \citealt{mccut95}; (8) \citealt{lada84};
(9) Distance discussed in \citealt{fvdt99}; (10) \citealt{lest86}; (11) \citealt{cram74}; (12) \citealt{wern79};
(13) \citealt{cram78}; (14) \citealt{cesar97}; (15) \citealt{chan93b}; (16) \citealt{turn84}; 
(17) \citealt{harv83}; (18) \citealt{neck78}; (19) \citealt{harv94}; (20) \citealt{acor98}; (21) \citealt{tief97}; 
(22) \citealt{simon83}; (23) \citealt{reng96}; (24) \citealt{camp84}; 
(25) \citealt{evan89}; (26) \citealt{prat92};
(27) \citealt{tofan95}; (28) \citealt{john84}; (29) \citealt{reid95}; 
(30) \citealt{rodr82}; (31) \citealt{wood89a}}

\end{deluxetable}

\pagebreak

\begin{deluxetable}{rrrrrrrrrr}
\tablewidth{0pt}
\tabletypesize{\scriptsize}
\tablecaption{Observed line fluxes $\int T_{\rm mb}dV$ (K~\kms).\label{t:lineflux}}

\tablehead{
\colhead{Line} &
\colhead{W~3}  & 
\colhead{GL~490}  & 
\colhead{W~33A} & 
\colhead{GL~2136} &
\colhead{S~140} & 
\colhead{NGC~7538} & 
\colhead{NGC~7538} &
\colhead{NGC~6334} &
\colhead{W~3} \\
 &
\colhead{IRS5} & & & &
\colhead{IRS1} & 
\colhead{IRS1} & 
\colhead{IRS9} & 
\colhead{IRS1} & 
\colhead{(H$_2$O)} 
       }
\startdata

\multicolumn{10}{c}{JCMT observations} \\
 & & & & & & & & & \\

C$^{17}$O 2--1
\tablenotemark{(a)}
                   &12.2 &\nodata &11.3&6.8 &6.4 &8.7 &3.8 &26.5&\nodata \\
 3--2              &16.6 &4.9 &15.7&9.9 &10.8&17.9&4.5 &57.9&15.1 \\
CS 5--4            &\nodata&16.5&36.2&11.2&30.2&54.8&20.6&\nodata &\nodata \\
 7--6              &25.2 & 8.0&17.4&4.7 &22.3&56.0&20.5&162.&83.7 \\
 10--9             &16.1 &\nodata &\nodata &\nodata &\nodata &34.4&3.3 &\nodata &78.4 \\
C$^{34}$S 5--4\tablenotemark{(b)}
              &3.7 &1.1 &5.6&1.1  &2.7 &7.0 &2.4 &56.5&20.7 \\
          7--6&2.5 &0.6 &5.3\tablenotemark{(d)}
                            &$<0.4$&2.1&7.9\tablenotemark{(d)}
                                           &3.1 &64.8\tablenotemark{(d)}
                                                     &25.8 \\
         10--9&2.2 &\nodata &\nodata &\nodata &\nodata &4.6 &$<3.4$&\nodata &31.6 \\
\hhco\ $3_{03}-2_{02}$&8.5 &4.4 &13.1& 4.4&13.7&27.0&14.2&63.2&20.2 \\
 $3_{12}-2_{11}$&19.0&\nodata &15.0& 6.3&19.2&27.0&13.6&63.0&41.2 \\
 $3_{22}-2_{21}$&3.6 &0.7 &3.5 & 1.6&3.5 &7.1 &3.7 &26.2&6.9 \\
 $5_{05}-4_{04}$&5.2 &\nodata &10.9& 4.3&8.7 &18.2&9.5 &82.9&31.1 \\
 $5_{15}-4_{14}$&14.0&\nodata &20.0& 7.4&20.5&28.8&\nodata &105.&36.2 \\
 $5_{24}-4_{23}$&2.7 &1.6 &4.5 & 2.2&3.7 &7.9 &4.5 &52.9&19.5 \\
 $5_{32}-4_{31}$&2.9 & 1.5 &5.4&1.5 &3.6 &8.8 &4.5 &48.1&33.4\tablenotemark{(e)} \\
 $5_{33}-4_{32}$&2.6 & 1.2 &4.9&2.6 &3.3 &9.9 &4.2 &50.6&24.7 \\
 $5_{42/41}-4_{41/40}$&0.9&$<0.7$&1.5&$<0.4$&$<0.5$&2.4&1.3&29.4&8.3 \\
 $7_{17}-6_{16}$&21.9&\nodata &\nodata &\nodata &7.3 &15.8&10.8&\nodata &24.3 \\
 & & & & & & & & & \\
\multicolumn{10}{c}{IRAM 30m observations} \\
 & & & & & & & & & \\
CS 5--4        &\nodata &\nodata &80.7    &22.5 &37.7 &79.0    &37.2 &\nodata &\nodata \\
C$^{34}$S 2--1 &\nodata &\nodata &7.6     &1.8  &3.6  &\nodata &4.1  &\nodata &\nodata \\
C$^{34}$S 3--2 &\nodata &\nodata &\nodata &2.4  &4.3  &\nodata &5.4  &\nodata &\nodata \\
 & & & & & & & & & \\
\multicolumn{10}{c}{CSO observations} \\
 & & & & & & & & & \\
CS 5--4\tablenotemark{(c)} 
        &21.5 &\nodata &20.3 &8.3 &30.5 &54.4 &18.2   &78.6  &33.5 \\
CS 7--6 &20.0 &4.9     &10.5 &6.6 &20.7 &50.7 &$<0.2$ &103.6 &64.2 \\
C$^{34}$S 7--6 &\nodata &\nodata &\nodata &\nodata &\nodata &\nodata &\nodata &\nodata &15.8 \\
\enddata

\tablenotetext{(a)}{For this line, a flux of 2.9 K~\kms\ was measured
  for IRAS~20126 and 24.5 K~\kms\ for DR~21~(OH).}

\tablenotetext{(b)}{For this line, a flux of 0.5 K~\kms\ was measured
  for GL~7009S,  3.2 K~\kms\ for IRAS~20126 and 57.9 K~\kms\ for W~28~A2.}

\tablenotetext{(c)}{For this line, a flux of $<0.8$ K~\kms\ was measured
  for GL~7009S,  $<0.6$ K~\kms\ for IRAS~20126, 59.7 K~\kms\ for DR~21~(OH) 
  and 123.2 K~\kms\ for W~28~A2.}

\tablenotetext{(d)}{Blend with triplet of CN lines; fit constrained by
  requiring same $\Delta V$ as other \cs\ and \co\ lines.}

\tablenotetext{(e)}{Blend with CH$_3$OCHO line complex; fit constrained by
  requiring same $\Delta V$ as other \hhco\ lines.}

\end{deluxetable}

\pagebreak

\begin{deluxetable}{lrrrr}
\tablewidth{0pt}
\tablecaption{Single-dish line emission: basic parameters.\label{vlsr.tab}}

\tablehead{
\colhead{Source} &    
\colhead{\vlsr}  &
\colhead{$\Delta V$} &
\colhead{N(CO)} &
\colhead{D(CS)}  \\
 &
\colhead{\kms} &
\colhead{\kms} &
\colhead{$10^{19}$~\scm} &
\colhead{arcsec} \\
}

\startdata
                                                        
W~3 IRS5        &   -38.4 (0.3) &  2.7 (1.1)  & 3.7  & 55 \\
GL 490          &   -13.3 (0.2) &  3.3 (0.6)  & 0.8  & 38 \\
W~33A           &   +37.5 (0.9) &  5.4 (0.5)  & 3.1  & 37 \\
GL 2136         &   +22.8 (0.1) &  3.1 (0.4)  & 1.4  & 36 \\
GL 7009S        &   +40.3 (0.6) &  5.0 (1.0)  & \nodata & \nodata  \\
GL 2591\tablenotemark{(a)}
                &    -5.5 (0.2) &  3.3 (0.6)  & 3.4  & 52 \\
S~140 IRS1      &    -7.0 (0.2) &  3.3 (0.3)  & 1.4  & 64 \\
NGC~7538 IRS1   &   -57.4 (0.5) &  4.1 (1.4)  & 3.9  & 52 \\
NGC~7538 IRS9   &   -57.2 (0.3) &  4.1 (0.5)  & 1.0  & 47 \\
                &               &             &      &  \\
IRAS 20126\tablenotemark{(b)}
                &    -3.7 (0.2) &  3.2 (0.2)  & 0.8  & 40  \\
DR 21 (OH)      &    -3.1 (0.2) &  4.5 (0.2)  & 15.6 & 56 \\
W~3 (\water)    &   -47.6 (0.6) &  5.8 (0.6)  & 8.5  & 41 \\
NGC~6334 IRS1   &    -7.4 (0.2) &  5.3 (0.3)  & 16.9 & 50 \\
W 28 A2\tablenotemark{(c)}
                &    +10        &  5.7        & \nodata & 38 \\

\enddata

\tablecomments{The CO column densities in column~4 are derived from
  C$^{17}$O emission assuming $^{16}$O/$^{17}$O=2500, refer to a
  $14''$ beam and have a 30\% calibration uncertainty. Column~5 lists
  the half-power diameters of the CS $J=5\to4$ emission mapped with
  the CSO.}

\tablenotetext{(a)}{Size of CS emission from data presented in \citet{carr95}}
\tablenotetext{(b)}{Size of CS emission from data presented in \citet{cesar97}}
\tablenotetext{(c)}{Velocity and line width taken from \citet{plum97}}

\end{deluxetable}

\begin{deluxetable}{lrrrrr}
\tablewidth{0pt}
\tablecaption{Results of OVRO observations.\label{posflux.tab}}

\tablehead{
\colhead{Source} &
\colhead{Frequency} & 
\colhead{RA (1950)} & 
\colhead{Dec (1950)}  & 
\colhead{Flux Density} &
\colhead{Beam Size} \\
 & 
\colhead{(GHz)} & 
 &
 &
\colhead{(mJy)} & 
\colhead{(arcsec)} 
          }
\startdata
GL 2136   & 86 & 18:19:36.62 (0.02) & -13:31:43.8 (0.3) & 61 & $3.0\times 2.8$ \\
W 33A MM1 & 86 & 18:11:44.22 (0.02) & -17:52:58.2 (0.2) & 24 & $3.3\times 2.7$ \\
 & 106 & & & 54 & $2.9\times 2.6$ \\
 & 233 & & & 190 & $2.0\times 1.6$ \\
W 33A MM2 & 86 & 18:11:43.91 (0.02) & -17:52:59.8 (0.2) & 38 & $3.3\times 2.7$ \\
 & 106 & & & 30 & $2.9\times 2.6$ \\
 & 233 & & & 140 & $2.0\times 1.6$ \\
NGC 7538 IRS1 & 86 & 23:11:36.64 (0.02) & +61:11:49.8 (0.2) & 1500 & $2.6\times 2.0$ \\
 & 107 & & & 1800 & $2.4\times 1.9$ \\
 & 115 & & & 2000 & $2.5\times 1.7$ \\
 & 230 & & & 3500 & $1.1\times 0.9$ \\
NGC 7538 IRS9 & 107& 23:11:52.90 (0.07) & +61:10:59.2 (0.5) & 43 & $2.4\times 2.0$ \\
 & 115 & & & 95 & $2.4\times 1.9$ \\
\enddata

\tablecomments{The flux densities have a calibration uncertainty of
  $\approx 30$\%.}

\end{deluxetable}

\pagebreak

\begin{deluxetable}{llllrrcccc}
\tablewidth{0pt}
\tablecaption{Parameters of Best Fit Models\label{t:m_gas}}
  
\tablehead{
\colhead{Source} &
\colhead{$r_0$} &
\colhead{$\alpha$} &
\colhead{$n_0$} &
\colhead{\mass} &
\colhead{$M_V$} &
\colhead{$\bar{T}$} &
\colhead{CO/H$_2$} &
\colhead{CS/H$_2$} &
\colhead{H$_2$CO/H$_2$} \\
 &
\colhead{$10^3$ AU} &
 &
\colhead{$10^4$ cm$^{-3}$} &
\colhead{\msol} &
\colhead{\msol} &
\colhead{K} &
\colhead{$\times 10^4$} &
\colhead{$\times 10^9$} &
\colhead{$\times 10^9$} }

\startdata
W~3 IRS5      &60 &1.5 &$2.3$&  262 & 355 &33 &1.6 &5. & 3.\\
GL 490        &19 &1.5 &$8.6$&   30 & 168 &19 &0.4 &1. & 1.\\
W~33A         &74 &1.0 &$6.8$& 1089 &1984 &20 &0.5 &5. & 4.\\
GL 2136       &35 &1.25&$3.6$&   72 & 316 &28 &1.2 &4. & 8.\\
GL 7009S      &91 &0.5 &$17.$& 3955 &2218 &21 &\nodata &0.4-0.8 & \nodata \\
GL 2591       &27 &1.0 &$5.8$&   44 & 268 &28 &1.5 &10.& 4.\\
S~140 IRS1    &29 &1.5 &$3.6$&   46 & 128 &27 &1.0 &5. & 5.\\
NGC~7538 IRS1 &72 &1.0 &$5.3$&  815 &1128 &25 &0.6 &10.&10.\\
NGC~7538 IRS9 &66 &1.0 &$3.9$&  430 &1016 &20 &0.3 &10.&10.\\
IRAS 20126    &39 &1.75&$2.2$&   84 & 294 &27 &0.5 &3.& \nodata \\
DR~21 (OH)    &29 &1.75&$23.$&  350 & 429 &15 &0.8 &3.& 5.\\
W~3 (H$_2$O)  &45 &2.0 &$5.3$&  385 & 932 &26 &1.0 &10.& 3.\\
NGC~6334 IRS1 &43 &2.0 &$6.2$&  382 & 725 &35 &1.5 &12.& 7.\\
W~28 A2       &37 &1.5 &$9.9$&  292 & 982 &42 &1.7 &20.& \nodata \\
\enddata

\tablecomments{The parameters in columns~2-4 specify the
  density structure in the form $n(r)=n_0 (r/r_0)^{-\alpha}$, with the
  reference radius $r_0$ half the diameter listed in Table~3. Column~5
  gives the integrated mass enclosed within $r_0$, and column~5 the
  virial mass within $r_0$.} 

\end{deluxetable}

\pagebreak

\end{document}